\documentclass[11pt,a4paper]{article}
\usepackage{jheppub}
\usepackage{tikz}
\usepackage{tikz-3dplot}
\allowdisplaybreaks

\usepackage{float}

\newcommand{\Res}{\mathop{\mathrm{Res}}\nolimits}

\title{Five-point partial waves, splitting constraints and hidden zeros}
\author{Arnab Priya Saha$^{a}$ and }
	\emailAdd{arnabsaha@iisc.ac.in}
	\author{Aninda Sinha$^{a,b}$}%
	\emailAdd{asinha@iisc.ac.in}
	\affiliation{%
		${}^a$Centre for High Energy Physics,
		Indian Institute of Science,\\
		C.V. Raman Avenue, Bangalore 560012, India.
	}
	\affiliation{%
		${}^b$Department of Physics and Astronomy, 
		University of Calgary,\\
		Alberta T2N 1N4, Canada
	}
\begin{document}
\abstract{
We introduce a partial-wave basis for the double residues of five-point tree amplitudes involving identical external scalar particles, decomposing them into exchanges of definite spin at each internal channel. We verify this basis using massive spinor-helicity building blocks and by matching the resulting partial-wave coefficients against the tree-level five-point Veneziano amplitude at fixed mass levels.
As an application, we express five-point splitting constraints --- the reduction of the five-point amplitude to products of four-point amplitudes on special kinematic loci --- as linear relations among the five-point partial-wave coefficients. At low mass levels these constraints, together with spin truncation, fix the full five-point partial-wave data in terms of the four-point coefficients and imply simple compatibility conditions; remarkably, imposing two independent splitting loci also forces the residue to vanish on their intersection, making the associated hidden zero manifest in partial-wave space. We also show that once both channels allow spin-2 exchange a genuine kernel can remain, indicating the need for additional higher-point input to achieve complete rigidity.

}

	\maketitle
    
	\flushbottom
	
	\section{Introduction}

    In recent times, the modern S-matrix bootstrap program has been used to study constraints arising from crossing symmetry, unitarity, high energy behaviour and dispersive sum rules---using certain extra input, one has been able to explore the region in the space of S-matrices where stringy physics lives \cite{Guerrieri:2021ivu, Kruczenski:2022,  deRham:2022gfe, Haring:2023, Cheung:2024Strings}.  However, most bootstrap technology and most quantitative bounds have so far been developed in the simplest setting of $2\!\to\!2$ scattering.  This makes it especially timely to systematically extend the program to five-particle (and, more generally, $n$-particle) amplitudes, where genuinely new physics and new constraints first appear: inelasticity and multi-particle unitarity, a richer network of factorization channels, and kinematic configurations that are invisible at four points.  Indeed, recent work has demonstrated that higher-point consistency can feed back nontrivially into lower-point data—for example, “hidden-zero’’ and “splitting’’ conditions \cite{Arkani-Hamed:2023swr} on five-point EFT amplitudes can drastically reduce the allowed region for four-point Wilson coefficients and drive the bootstrap toward the open-string beta function \cite{Berman:2025splittingregions}.  More broadly, mixed-multiplicity positivity constraints (“multipositivity’’) provide an infinite family of inequalities that simultaneously constrain residues and EFT coefficients at arbitrary multiplicity, and are strikingly saturated by tree-level open-string amplitudes \cite{Cheung:2025Multipositivity}.  Complementing these S-matrix developments, multipoint bootstrap methods on the CFT side—such as positive semi-definite reformulations of six-point-crossing underscore that going beyond four points is both feasible (but technically challenging) and generically more constraining, e.g. \cite{Bercini:2020msp, poland, apratim5}.  Finally, rapid advances in multi-leg amplitude technology and function-space control---including powerful new methods for evaluating and organizing multi-point integrals---make five- and six-particle kinematics increasingly tractable, strengthening the case for a systematic study of higher-point amplitudes and their analytic structures and factorization properties \cite{Henn:2024SixPointIntegrals}.

Extending S-matrix ideas beyond $2\!\to\!2$ scattering is essential but technically formidable. Even at tree-level, knowledge of multipoint amplitude can help extract the underlying vertices and construct loop amplitudes perturbatively. At least in principle, this points at a systematic examination of perturbative unitarity of amplitudes originating from tree-level S-matrix bootstrap.  Multiparticle amplitudes are expected to live on a far more intricate multi-sheeted Riemann surface than four-point scattering, with many independent invariants constrained by Gram-determinant relations and with a proliferation of normal and anomalous thresholds \cite{Eden:1966Book}.  In the late 1960s, five-point amplitudes became a central testing ground:  requirement of duality  and Regge behavior led to explicit    five-point constructions  and generalizations of Veneziano/Virasoro-type formulae \cite{Virasoro:1969, Mandelstam:1969}, while multi-Regge theory led to a detailed framework for studying analyticity and factorization of the five-point function \cite{Abarbanel:1972ayr, White:1973, Brower:1974}, mostly in certain limits.  A major obstruction to a systematic bootstrap at higher multiplicity is that analyticity constraints are no longer captured by a single complex variable: different channel discontinuities can overlap, and consistency requires nontrivial Steinmann relations, which dictate that double discontinuities on overlapping channels vanish, together with generalized optical theorems expressing multiparticle cuts in terms of physical-region amplitudes \cite{Steinmann:1960, Cahill:1973}.  In practice, these features have made it difficult to write higher-point analogs of the machinery, such as dispersive representations or partial-wave expansions, which are conventionally used to obtain bounds in $2\!\to\!2$ scatterings \cite{deRham:2017avq,  deRham:2017zjm, Tolley:2020gtv, Caron-Huot:2020cmc,  Sinha:2020win}.

Even setting aside analyticity, the most basic remaining challenge is unitarity: beyond four points, unitarity becomes genuinely nonlinear and couples different multiplicities through phase-space integrals for loops, so imposing it sharply in a bootstrap-friendly way remains difficult.  An interesting new route is provided by “hidden zeros’’—special kinematic loci where higher-point amplitudes vanish with a factorization-like structure—because these conditions translate nontrivial higher-point consistency into comparatively tractable algebraic constraints that can be fed into EFT/bootstraps, dramatically shrinking allowed regions and isolating string-like solutions \cite{Bartsch:2024HiddenZeros, Berman:2025splittingregions}.

    The  remarkable discovery of ``hidden zeroes" in amplitudes \cite{Arkani-Hamed:2023swr} begins with the observation that certain color-ordered massless tree-amplitudes vanish on a highly non-generic locus in kinematic space, defined by setting a special set of \emph{non-planar} Mandelstam invariants to zero. In \cite{Arkani-Hamed:2023swr} these zeros are not presented as an accident of Feynman-diagram numerators; rather, they become manifest in the positive geometric description of the amplitudes using canonical forms \cite{Arkani-Hamed:2017mur}.  Even more striking is what happens \emph{near} the zero locus: the amplitude exhibits a novel ``(2-)split'' behavior, simplifying into a nontrivial product of lower-point objects (up to a known kinematic prefactor) without being on an ordinary factorization pole.  The same zero locus and splitting structure is shown to extend beyond $\mathrm{Tr}(\phi^{3})$ to pions (NLSM) and non-supersymmetric gluons, and a unique kinematic shift preserving the zeros relates these theories as different ``faces'' of a single underlying function; the phenomenon further generalizes to ``stringy'' $\mathrm{Tr}(\phi^{3})$ amplitudes \cite{Arkani-Hamed:2023swr}.

The hidden zeroes and splitting are considered from a \emph{bootstrap input} point of view with sharp EFT consequences in \cite{Berman:2025splittingregions}.  Working within the perturbative numerical $S$-matrix bootstrap for weakly coupled EFTs, this work imposes the hidden-zero conditions and, crucially, the \emph{five-point} splitting constraints on a systematic EFT ansatz.  While the hidden-zero conditions largely fix five-point contact ambiguities, enforcing the splitting relations propagates back to $2\!\to\!2$ scattering by generating \emph{nonlinear} constraints among the four-point Wilson coefficients \cite{Berman:2025splittingregions}.  When incorporated into positivity-based bootstrap analyses, these higher-point constraints can shrink and deform the allowed regions for the four-point EFT data into smaller non-convex islands.

In this paper we develop a convenient partial-wave language for \emph{planar-ordered} tree-level five-point amplitudes, with ${\rm tr} \left(\phi^3\right)$ theory and the tree-level five-point Veneziano amplitude \cite{Arkani-Hamed:2024nzc} as guiding examples.  
Despite the central role of partial-wave methods in $2\!\to\!2$ bootstrap bounds, an equally practical and widely adopted partial-wave technology for five-point, \textit{i.e.,} $2\rightarrow 3$ amplitudes has not been discussed in the modern literature.  The reason is partly kinematic and partly analytic: five-point scattering depends on several independent invariants subjected to Gram-determinant constraints, and the amplitude lives on a considerably richer multi-sheeted analytic structure with overlapping discontinuities and multiparticle thresholds, so more than single angular variable are required to describe the five-point kinematic. Higher point parameterization of kinematic variables using little group frames can be found in \cite{Toller:1969gx}.  As a result, most recent higher-point bootstrap discussions have emphasized on polynomial residue ans\"atze and consistency conditions (such as hidden zeros and splitting) rather than a systematic harmonic decomposition.

We believe that this leaves a useful gap.  Our work supplies a concrete partial-wave basis and inversion formula for planar five-point double residues, validated by massive spinor-helicity analysis in four dimensions. The Veneziano amplitude serves as a testing ground for the five-point basis. This also provides a natural language in which higher-point constraints (including splitting/hidden-zero conditions) become transparent linear relations on partial-wave data.  In this sense, our framework is intended to bring to five points a level of partial-wave control that has long been standard at four points.

Working with the independent Mandelstam invariants $(s_{12},s_{23},s_{34},s_{45},s_{51})$, we focus on double residues on a compatible pair of factorization channels, e.g.   poles at $s_{12}$ and $s_{34}$.  We show that the remaining kinematics can be parameterized by three angles $(\theta_{12},\theta_{34},\omega)$, and we develop a $d$-dimensional harmonic analysis of the residues on this space coordinatized by the three angles.  This leads to an explicit basis expansion for double residues in terms of Gegenbauer polynomials (which reduces to associated Legendre polynomials in $d=4$) having cosines of $\theta_{12}$ and $\theta_{34}$ in the arguments together with Fourier modes $e^{in\omega}$. We also derive a practical inversion formula (\ref{pwcoeff-d}) for the partial-wave coefficients $a^{(s_{12}, s_{34})}_{jln}$ entering the absorptive part of the amplitude.

In four dimensions, we further validate and interpret the construction by gluing three-point amplitudes with massive higher-spin exchange using spinor-helicity variables, which fixes the allowed angular structures and, in particular cases, the relative weights of the $\omega$-harmonics.  We then apply this framework to the Veneziano amplitude as a nontrivial  check and as a source of exact data for the coefficients.  Finally, we translate five-point splitting conditions on special kinematic loci into linear constraints on the five-point partial-wave coefficients.  In the examples where spins of the exchanges at any channel is below 2, imposing two independent splitting conditions together with spin truncation fixes the full five-point partial-wave data in terms of four-point partial waves and enforces simple compatibility conditions on the latter. In cases where both channels have exchanges of spin 2 and higher spin states, we find a genuine residual kernel, suggesting that additional higher-point input is required to achieve complete rigidity. However, quite remarkably, we find that at the level of the residues of the five-point amplitude, the splitting identities imply the vanishing of the residues on the intersection --- in other words it points at hidden zeros for a larger class of Veneziano-like amplitudes.

Complementary progress on higher-point constraints comes from the recent ``multipositivity'' program of \cite{Cheung:2025Multipositivity}.  Their approach does not rely on a five-point partial-wave decomposition; instead, it organizes tree amplitudes by residues in multi-particle (half-ladder) factorization channels and shows that these residues can be interpreted as Gram matrices/inner products of couplings.  Positivity of these Gram matrices implies an infinite hierarchy of mixed-multiplicity inequalities (schematically relating $4$-, $5$-, $6$-point data, e.g.\ of Cauchy--Schwarz type), which translate into strong constraints on low-energy EFT coefficients and are strikingly saturated by open-string tree amplitudes \cite{Cheung:2025Multipositivity}.  Our perspective is complementary: we develop a harmonic (partial-wave) basis for five-point residues and use it to express splitting/hidden-zero constraints as linear relations among five-point partial-wave coefficients and induced compatibility conditions on four-point data.  Our work is also complementary to \cite{Berman:2025splittingregions} in two respects.  First, we provide a systematic partial-wave language for planar five-point residues --- including explicit inversion formulas and spinor-helicity checks --- which makes the implementation and interpretation of higher-point constraints more transparent.  Second, we recast splitting constraints as linear relations among five-point partial-wave coefficients (together with compatibility conditions on four-point partial waves), and we illustrate explicitly how and when these relations fix the five-point data (and when a genuine kernel remains once higher spins are allowed), thereby supplying an analytic partial-wave viewpoint on the rigidity underlying ``shrinking islands''.

\paragraph{Summary of results.}
Our main results can be summarized as follows.
\begin{itemize}
\item We parameterize the kinematics of five-point double residues on a compatible pair of poles (e.g.\ $s_{12}=m^2+k_1$ and $s_{34}=m^2+k_2$ for positive integer values of $k_{1}$ and $k_{2}$) by three angles $(\theta_{12},\theta_{34},\omega)$ and develop a $d$-dimensional harmonic basis (Gegenbauer polynomials and Fourier modes) together with an explicit inversion formula for the coefficients $a^{(k_1,k_2)}_{jln}$.
\item In $d=4$ we validate and interpret the basis using massive spinor-helicity gluing of three-point amplitudes with higher-spin exchange, including a particularly sharp simplification for equal-mass exchange where only the $j=l$ sector survives and the relative $\omega$-harmonic weights are fixed.
\item Using the tree-level five-point Veneziano amplitude as an exact testbed, we match residues at fixed mass levels and extract explicit partial-wave data.
\item We translate five-point splitting/hidden-zero factorization loci into linear constraints on the five-point partial-wave coefficients and induced compatibility conditions on four-point data; at low levels these constraints can fix the full five-point residue, while once spin-2 exchange is allowed for both exchanges a genuine kernel can remain, pointing to the need for additional higher-point input.
\end{itemize}

This paper is organized in the following manner: In Sec.(\ref{sec:fivepoint}) we present the kinematic parameterizations of the five-point amplitudes and introduce the basis for partial-wave expansions. Calculation details are relegated to Appendices (\ref{app:5ptparameter}) and (\ref{App:MultiRegge}) and a brief review of four-point kinematics is provided in Appendix (\ref{App:4-pt}). Checks of the basis in four dimension, done using massive spinor-helicity variables are given in Sec.(\ref{Sec:5pt-SH}). Transformation rules for the spinor-helicity variables can be found in Sec.(\ref{App:SH-param}). In Sec.(\ref{Sec:5pt-Constr}) we analyze the constraints on five-point coefficients with the algebraic details put in Appendix (\ref{app:splitting-algebra}). We conclude with a discussion of future directions.

\section{Five-point amplitudes} \label{sec:fivepoint}
	A five-point amplitude has five independent Mandelstam variables. Since we are interested in planar-ordered amplitudes, it will be convenient for us if we choose the five variables as  
	\begin{equation}
		s_{12}~, \quad s_{23}~, \quad s_{34}~, \quad s_{45}~, \quad s_{51}~.
	\end{equation}
    Our conventions are $s_{ij}=-\left(p_{i}+p_{j}\right)^{2}$ and $p_{i}^{2}=-m^{2}$. 
	 The simplest amplitude that can be written with the above variables is
	\begin{eqnarray}\label{phi3}
		\mathcal{M}^{\phi^3}_{5} &=& \frac{1}{\left(s_{12}-m^{2}\right)\left(s_{34}-m^{2}\right)} + \frac{1}{\left(s_{23}-m^{2}\right)\left(s_{45}-m^{2}\right)} + \frac{1}{\left(s_{34}-m^{2}\right)\left(s_{51}-m^{2}\right)} \nonumber\\
		&& + \frac{1}{\left(s_{45}-m^{2}\right)\left(s_{12}-m^{2}\right)} + \frac{1}{\left(s_{51}-m^{2}\right)\left(s_{23}-m^{2}\right)}~,
	\end{eqnarray} 
    which corresponds to $\text{tr}\left(\phi^{3}\right)$ theory. The trace is taken over some putative color group.  Note that this amplitude remains invariant under simultaneous exchanges of the following variables 
	\begin{equation}\label{cs-5pt}
		s_{12}\leftrightarrow s_{34} \quad \& \quad s_{45}\leftrightarrow s_{51}~,
	\end{equation}
	along with their cyclic permutations. There are five such transformations possible.
    
    Another non-trivial example of amplitude which is preserved under exchanges given in \eqref{cs-5pt} is the five-point Veneziano amplitude, a closed form expression of which is recently derived in \cite{Arkani-Hamed:2024nzc},
	\begin{eqnarray}\label{Ven-5pt}
		\mathcal{M}^{\text{Ven}}_{5}&=&\Gamma \left(m^2-s_{12}\right) \Gamma \left(m^2-s_{23}\right) \Gamma
		\left(m^2-s_{34}\right) \Gamma \left(m^2-s_{45}\right) \Gamma \left(m^2-s_{51}\right)\\
		&&
		\Gamma \left(\frac{1}{2} \left(3 m^2-s_{34}-s_{45}-s_{51}-1\right)\right) \Gamma
		\left(3 m^2-s_{34}-s_{45}-s_{51}\right) \nonumber\\
		&& \,
		_6\tilde{F}_5\left( \begin{matrix}
			m^2-s_{51}\\
			m^2-s_{34}\\
			m^2+s_{12}-s_{34}-s_{45}\\
			3
			m^2-s_{34}-s_{45}-s_{51}-1\\
			\frac{1}{2} \left(3
			m^2-s_{34}-s_{45}-s_{51}+1\right)\\
			m^2+s_{23}-s_{45}-s_{51}
		\end{matrix}; \begin{matrix}
			2 m^2-s_{12}-s_{51}\\
			2
			m^2-s_{45}-s_{51}\\
			2 m^2-s_{34}-s_{45}\\
			\frac{1}{2} \left(3
			m^2-s_{34}-s_{45}-s_{51}-1\right)\\
			2 m^2-s_{23}-s_{34}
		\end{matrix} ;-1\right)~.\nonumber
	\end{eqnarray}	
    We have used the tilde to denote regularized hypergeometric function, $_p\tilde{F}_{q}\left(a_{1},\ldots,a_{p};b_{1},\ldots,b_{q};z\right)=\frac{_{p} F_{q}\left(a_{1},\ldots,a_{p};b_{1},\ldots,b_{q};z\right)}{\Gamma\left(b_{1}\right)\ldots \Gamma\left(b_{q}\right)}$.
	Residue of this amplitude on the lowest mass-level of any factorization channel is the four-point Veneziano amplitude. For example, the residue on $s_{34}=m^{2}$ is 
	\begin{equation}
		\text{Res}_{s_{34}=m^{2}}\mathcal{M}^{\text{Ven}}_{5} = -\frac{\Gamma\left(m^{2}-s_{12}\right)\Gamma\left(m^{2}-s_{51}\right)}{\Gamma\left(2m^{2}-s_{12}-s_{51}\right)}~.
	\end{equation}
	More generally, residue on the compatible pair of factorization channels is a polynomial of finite degree in the other three kinematic variables. This polynomial contains information about spectrum of the exchanged states at any particular mass-level. Note that the arguments appearing inside the $\Gamma$ functions are dimensionless quantities - kinematic variables, $s_{ij}$ and the external mass, $m$ should be thought of as being scaled by appropriate powers of a mass parameter, $M$. For the rest of the discussion, unless mentioned otherwise, we will assume $M=1$.
	
	\subsection{Parameterization of kinematic variables}
	Let us consider a specific pair of factorization channels, $s_{12}$ and $s_{34}$, for any  planar-ordered five-point tree-amplitude.  The residue will be  a polynomial function of $s_{23}$, $s_{45}$ and $s_{51}$. These three kinematic variables can be expressed in terms of three angle parameters as shown in \cite{White:1971fz,White:1972sc}, 
	\begin{eqnarray}\label{kinem-param}
		s_{51} & = & \frac{1}{2}\left[3m^{2}-s_{12}+s_{34}+\sqrt{\frac{s_{12}-4m^{2}}{s_{12}}\biggl\{\left(s_{12}-s_{34}\right)^{2}-2m^{2}\left(s_{12}+s_{34}\right)+m^{4}\biggr\}}\cos\theta_{12}\right]~,\nonumber\\
		s_{45} & = & \frac{1}{2}\left[3m^{2}+s_{12}-s_{34}-\sqrt{\frac{s_{34}-4m^{2}}{s_{34}}\biggl\{\left(s_{12}-s_{34}\right)^{2}-2m^{2}\left(s_{12}+s_{34}\right)+m^{4}\biggr\}}\cos\theta_{34}\right]~,\nonumber\\
		s_{23} & = & \frac{1}{4}\left[\sqrt{\left(s_{12}-s_{34}\right)^{2}-2m^{2}\left(s_{12}+s_{34}\right)+m^{4}}\Biggl\{\sqrt{\frac{s_{12}-4m^{2}}{s_{12}}}\cos\theta_{12}-\sqrt{\frac{s_{34}-4m^{2}}{s_{34}}}\cos\theta_{34}\Biggr\}\right.\nonumber\\
		&& \left. +\sqrt{\frac{\left(s_{12}-4m^{2}\right)\left(s_{34}-4m^{2}\right)}{s_{12}s_{34}}}\left(s_{12}+s_{34}-m^{2}\right)\cos\theta_{12}\cos\theta_{34}\right.\nonumber\\
		&&\left.  -2\sqrt{\left(s_{12}-4m^{2}\right)\left(s_{34}-4m^{2}\right)}\sin\theta_{12}\sin\theta_{34}\cos\omega -s_{12}-s_{34}+9m^{2}\right]~.
	\end{eqnarray}
	The derivation of the above equations is given in appendix \ref{app:5ptparameter} along with an interpretation of the angles. Under simultaneous exchanges of 
    \begin{equation}
        s_{12} \leftrightarrow s_{34} \quad \& \quad \theta_{12} \leftrightarrow \pi-\theta_{34}
    \end{equation}
    it can be seen that $s_{45}$ and $s_{51}$ are interchanged while $s_{23}$ remains unchanged. This is reminiscent of the symmetry alluded to in \eqref{cs-5pt}.

    \subsection{A compact form of the kinematic relations for $s_{51},\,s_{45},\,s_{23}$}

It is useful to rewrite the relations in \eqref{kinem-param} by packaging the nested square-roots into the standard K\"all\'en function and ``velocity'' factors.  Define
\begin{align}
\lambda \equiv \lambda(s_{12},s_{34},m^2)
&:= (s_{12}-s_{34})^2-2m^2(s_{12}+s_{34})+m^4
\nonumber\\
&= (s_{12}+s_{34}-m^2)^2-4s_{12}s_{34},\qquad
\Delta:=\sqrt{\lambda},
\label{eq:KallenDelta}
\\[2pt]
\beta_{12}:=\sqrt{1-\frac{4m^2}{s_{12}}},\qquad
&\beta_{34}:=\sqrt{1-\frac{4m^2}{s_{34}}}.
\label{eq:betadef}
\end{align}
Then the first two relations in eq.~(2.6) become affine in $\cos\theta$:
\begin{align}
s_{51}&=\frac{3m^2-s_{12}+s_{34}}{2}+\frac{\Delta}{2}\,\beta_{12}\cos\theta_{12},
\label{eq:s51compact}\\
s_{45}&=\frac{3m^2+s_{12}-s_{34}}{2}-\frac{\Delta}{2}\,\beta_{34}\cos\theta_{34}.
\label{eq:s45compact}
\end{align}
Moreover, the expression for $s_{23}$ can be written as
\begin{align}
s_{23} &= \frac{9m^2-s_{12}-s_{34}}{4}
+\frac{\Delta}{4}\Big(\beta_{12}\cos\theta_{12}-\beta_{34}\cos\theta_{34}\Big)
\nonumber\\[3pt]
&\quad+\frac{\beta_{12}\beta_{34}}{4}(s_{12}+s_{34}-m^2)\cos\theta_{12}\cos\theta_{34}
-\frac{1}{2}\sqrt{s_{12}s_{34}}\,\beta_{12}\beta_{34}\,\sin\theta_{12}\sin\theta_{34}\cos\omega.
\label{eq:s23compact}
\end{align}
Finally, using \eqref{eq:s51compact}--\eqref{eq:s45compact} one may eliminate the combination
$\Delta(\beta_{12}\cos\theta_{12}-\beta_{34}\cos\theta_{34})$ in favour of invariants:
\begin{equation}
\Delta\big(\beta_{12}\cos\theta_{12}-\beta_{34}\cos\theta_{34}\big)
=2\big(s_{51}+s_{45}\big)-6m^2,
\label{eq:DeltaLinearElim}
\end{equation}
which can be convenient when trading angular variables for Mandelstam invariants.

	\subsection{Partial-wave coefficients for residues}
	Partial-wave expansions of four-point amplitudes are well known in the literature. In general, for external states with arbitrary spins, the amplitudes can be expanded in terms of Wigner's D-matrix. If the external states are identical scalars, then the basis reduces to Gegenbauer polynomials in general dimension. There have been studies of partial-wave expansions for amplitudes beyond four-point \cite{White:1971fz, White:1972sc, White:1973, Abarbanel:1972ayr}. However, analytic structures of higher-point amplitudes are complicated and therefore in most of the cases the corresponding expansions have limited domains of convergence. For example, in \cite{White:1972sc, White:1973} partial-wave expansion of five-point amplitude in the combined $s_{12}$ and $s_{34}$ channels have been given in terms of the angle variables, $\theta_{12}$, $\theta_{34}$ and $\omega$ in \eqref{kinem-param}, 
    \begin{equation}
        \mathcal{M}_{5}\left(s_{12}, s_{23}, s_{34}, s_{45}, s_{51}\right) = \sum_{j=0}^{\infty}\sum_{l=0}^{\infty}\sum_{|n|\le j,l} f_{jln}\left(s_{12}, s_{34}\right)P_{j}^{-|n|}\left(\cos\theta_{12}\right) P_{l}^{-|n|}\left(\cos\theta_{34}\right)e^{i\omega n}~.
    \end{equation}
    \noindent\textbf{Remark.} In this paper we apply this expansion only to \emph{double residues} (fixed-level pole residues) of tree-level five-point amplitudes, which are polynomials in the remaining invariants and hence admit a finite spin truncation. We do not attempt a global partial-wave expansion of the full amplitude, whose singularities in other channels obstruct convergence.

    It can be seen that for certain values of cosines of the angles between $-1$ and $1$, either one or more of the kinematic variables - $s_{23}$, $s_{45}$ and $s_{51}$  can become positive. Since the amplitude generally has singularities in the positive domain of the kinematic variables, so the partial-waves inherit additional singularities, which come from the channels other than $s_{12}$ and $s_{34}$. In this work, we will restrict to five-point tree-level amplitudes which have only simple poles. The residues on the poles of any two compatible poles are polynomials in the other three variables. As the residues do not have any singularities, therefore we can always expand them in a suitable basis.  
    
   Motivated by the partial-wave expansion in \cite{White:1972sc}, we can write down a basis decomposition for the residues on the sets of poles at $s_{12}$ and $s_{34}$  of a five-point in general $d$ dimension,
	\begin{eqnarray}\label{pwexp-d}
		\text{Res}_{(s_{12}, s_{34})}\mathcal{M}_{5}
		& = &  \sum_{j,l}\sum_{n=-\min\{|j|,|l|\}}^{\min\{|j|,|l|\}} a_{jln}^{(k_{1},k_{2})}\left(\frac{2^{|n|}\Gamma\left(\frac{d-3}{2}+|n|\right)}{\sqrt{\pi}}\right)^{2} e^{i\omega n}\nonumber\\
		&& \times \left(1-x^{2}\right)^{\frac{|n|}{2}}\mathcal{G}_{j-|n|}^{\frac{d-3}{2}+|n|}\left(x\right)\left(1-y^{2}\right)^{\frac{|n|}{2}}\mathcal{G}_{l-|n|}^{\frac{d-3}{2}+|n|}\left(y\right)~.
	\end{eqnarray}
	We have defined $x=\cos\theta_{12}$ and $y=\cos\left(\pi-\theta_{34}\right)$. $j$ and $l$ denote spins of exchanged states in $s_{12}$ and $s_{34}$ channels respectively. This choice of basis can be understood from higher dimensional spherical harmonics \cite{2019arXiv190106711S}. $N$-dimensional spherical harmonics can be given by $Y_{l_{1},l_{2},\ldots,l_{N}}\left(
    \theta_{1}, \theta_{2}, \ldots, \theta_{N}\right) = \prod_{k=2}^{N}\left(\sin\theta_{k}\right)^{l_{k-1}}\mathcal{G}^{l_{k-1}+\frac{1}{2}\left(k-1\right)}_{l_{k}-l_{k-1}}\left(\cos\theta_{k}\right)e^{\pm i l_{1}\theta_{1}}$, where $0\le\theta_{1}<2\pi$, $0\le\theta_{i}\le\pi$ with $i=2,3,\ldots, N$ and $l_{N}\ge l_{N-1}\ge \ldots \ge l_{2}\ge l_{1} = |n|$ are the eigenvalues. In our case, little group for each of the massive states propagating in $s_{12}$ and $s_{34}$ channels is $SO\left(d-1\right)$, so $N=d-2$. We can orient the little group frame of each exchanged state along its $z$-axis. Due to symmetry, we can fix $d-4$ angles and the relevant transformations include the azimuthal angle and the angle subtended by the $z$-axis, which is either $\theta_{12}$ or $\theta_{34}$. The resultant angle formed by the azimuthal angles of the two frames is then identified with $\omega$, which is also called the Toller angle \cite{Brower:1974}: more details can be found in the appendices.  
    
	Using the orthogonality relation of Gegenbauer polynomials we can find the $d$ dimensional partial wave coefficients,
	\begin{eqnarray}\label{pwcoeff-d}
		a_{jln}^{(s_{12}, s_{34})} & = & \mathcal{N}_{d}^{-1}\int_{0}^{2\pi}\mathrm{d}\omega\int_{-1}^{1}\mathrm{d}x\int_{-1}^{1}\mathrm{d}y\; e^{-i\omega n}\left(1-x^{2}\right)^{\frac{d-4+|n|}{2}}\left(1-y^{2}\right)^{\frac{d-4+|n|}{2}}\nonumber\\
		&& \times \:  \mathcal{G}_{j-|n|}^{\frac{d-3}{2}+|n|}\left(x\right)\mathcal{G}_{l-|n|}^{\frac{d-3}{2}+|n|}\left(y\right)\text{Res}_{(s_{12}, s_{34})}\mathcal{M}_{5} ~,
	\end{eqnarray} 
	with the normalization factor is given by 
	\begin{equation}
		\mathcal{N}_{d} = 2\pi^{2}\frac{2^{-2\left(d-4+|n|\right)}}{\left[\Gamma\left(\frac{d-3}{2}+|n|\right)\right]^{2}}\frac{\Gamma\left(d-3+j+|n|\right)\Gamma\left(d-3+l+|n|\right)}{\left(\frac{d-3}{2}+j\right)\left(\frac{d-3}{2}+l\right)\Gamma\left(j-|n|+1\right)\Gamma\left(l-|n|+1\right)}~.
	\end{equation}
     One way to verify the expansion in \eqref{pwexp-d} is by gluing the three-point on-shell amplitudes on the poles of the factorization channels.  In four-dimension three-point amplitudes with massive higher spin states are classified using spinor-helicity variables \cite{Arkani-Hamed:2017jhn}. In the next section, we will work out some examples to illustrate the gluing procedure.  
	
	\paragraph{Coefficients in four dimension:} Plugging $d=4$ in \eqref{pwexp-d}, we obtain the basis expansion of the five-point residues in terms of associated Legendre polynomials\footnote{This differs from the basis in \cite{White:1972sc} in the choice of the sign of $|n|$ appearing in the associated Legendre polynomials. For generalization to higher dimension, we have used 
    \begin{equation}
	P_{j}^{n}\left(x\right) = \frac{\left(-2\right)^{n}\Gamma\left(n+\frac{1}{2}\right)}{\sqrt{\pi}}\left(1-x^{2}\right)^{\frac{n}{2}}\mathcal{G}_{j-n}^{n+\frac{1}{2}}\left(x\right)~.
\end{equation}
If we use negative sign for $|n|$, then the measure of integration in the inversion relation for finding partial-wave coefficients contains $\left(1-x^{2}\right)^{\frac{d-4-|n|}{2}}$. This factor gives potential divergences from the end points of integration in lower dimensions for higher values of spin.},
	\begin{equation}\label{pwexp-4}
		\text{Res}_{(s_{12}, s_{34})}\mathcal{M}_{5} = \sum_{j,l}\sum_{n=-\min\{|j|,|l|\}}^{\min\{|j|,|l|\}} a_{jln}^{(s_{12}, s_{34})}e^{i\omega n}P_{j}^{|n|}\left(x\right)P_{l}^{|n|}\left(y\right)~.
	\end{equation}
	Partial wave coefficients can be obtained by the inversion relation,
	\begin{eqnarray}
		a_{jln}^{(s_{12}, s_{34})} & = & \mathcal{N}_{4}^{-1}\int_{0}^{2\pi}\mathrm{d}\omega\int_{-1}^{1}\mathrm{d}x\int_{-1}^{1}\mathrm{d}y\; e^{-i\omega n}P_{j}^{|n|}\left(x\right)P_{l}^{|n|}\left(y\right)\text{Res}_{(s_{12}, s_{34})}\mathcal{M}_{5}~,
	\end{eqnarray}
	where the four-dimensional normalization factor is
	\begin{equation}
		\mathcal{N}_{4} = 8\pi\frac{\Gamma\left(j+|n|+1\right)}{\left(2j+1\right)\Gamma\left(j-|n|+1\right)}\frac{\Gamma\left(l+|n|+1\right)}{\left(2l+1\right)\Gamma\left(l-|n|+1\right)}~.
	\end{equation}

	\section{Massive spinor helicity checks in four dimensions }
    \label{Sec:5pt-SH}
	In this section we present a few examples of five-point residues with massive higher spin exchanges in the intermediate channels, which are $s_{12}$ and $s_{34}$. These residues are obtained by gluing three three-point sub-amplitudes using spinor-helicity variables \cite{Elvang:2013cua, Arkani-Hamed:2017jhn, Liu:2020fgu, KNBalasubramanian:2022sae}. This also helps us to determine the concerned partial wave coefficients in terms of the coupling constants of three-point amplitudes. As a result, this also serves as a check for the validity of \eqref{pwexp-4}. For simplicity we restrict to massless scalars in the external states. Parameterization of the spinor-helicity variables in terms of the angular variables are given in \ref{app:sphl-param}. We leave the analysis involving massive external states in the spinor-helicity formalism for future study. In this section all the Mandelstam variables will be considered to be dimensionful quantities. As opposed to the previous section, here $y=\cos\theta_{34}$ because we have already taken into account $\theta_{34}\rightarrow\pi-\theta_{34}$ while applying Lorentz transformations to the spinor-helicity variables.

	\subsection{Unequal mass exchanges}
	Let the mass and spin of the exchanged particle in $s_{12}$ channel be $m_{1}$ and $j$ respectively and that in the $s_{34}$ channel are $m_{2}$ and $l$. In this case, the middle sub-amplitude with one massless and two massive legs has $2\times \min\{j,l\}+1$ structures, each with a particular coupling constant. Three-point sub-amplitudes are classified in \cite{Arkani-Hamed:2017jhn}. We consider the following cases below. As for the notation, we will denote the massive state corresponding to $s_{12}$ channel as $P_{1}$ and that for $s_{34}$ channel as $P_{2}$.
	
	\paragraph{$j=1$ and $l=0$ :} The three-point sub-amplitudes are the following,
	\begin{eqnarray}
		\mathcal{A}_{1}\left(1,2,P_{1}^{(I_{1}I_{2})}\right) & = & \frac{g_{1}}{m_{1}}\langle 1P_{1}^{(I_{1}}\rangle\langle 2P_{1}^{I_{2})}\rangle[12]~,\nonumber\\
		\mathcal{A}_{2}\left(5,P_{1(I_{1}I_{2})},P_{2}\right) & = & g_{2}\langle P_{1(I_{1}}5\rangle[ P_{1 I_{2})}5]~,\nonumber\\
		\mathcal{A}_{3}\left(3,4,P_{2}\right) & = & g_{3}m_{2}~.
	\end{eqnarray}
	Product of these sub-amplitudes takes the form
	\begin{equation}
		\mathcal{A}_{1}\mathcal{A}_{2}\mathcal{A}_{3} = g_{1}g_{2}g_{3}\frac{m_{2}}{m_{1}^{2}}\langle1|2|1]^{2}\langle5|\left(p_{2}-p_{1}\right)|5]~.
	\end{equation}
	Applying the Lorentz transformations given in \eqref{spinor-transf} on the spinor-helicity variables, we obtain
	\begin{equation}
		\mathcal{A}_{1}\mathcal{A}_{2}\mathcal{A}_{3} = -g_{1}g_{2}g_{3}\frac{m_{2}}{m_{1}^{2}}s_{12}^{2}\left(s_{12}-s_{34}\right)\cos\theta_{12}~.
	\end{equation}
	This expression is proportional to $P_{1}^{0}\left(x\right)$.  It can be checked that under the interchange of spins of the exchanged particles, \textit{i.e.} $j=0$ and $l=1$, the product of the sub-amplitudes becomes 
	\begin{equation}
		\left(\mathcal{A}_{1}\mathcal{A}_{2}\mathcal{A}_{3}\right)_{j\leftrightarrow l} = -g_{1}g_{2}g_{3}\frac{m_{1}}{m_{2}^{2}}s_{34}^{2}\left(s_{34}-s_{12}\right)\cos\theta_{34}~,
	\end{equation}
	which is proportional to $P_{1}^{0}\left(y\right)$. This ensures that coefficient of the five-point residue remains the same under interchange of the exchanged states in the two factorization channels.
	
	\paragraph{$j=1$ and $l=1$ :} There are three possible ways how a massless scalar can couple to two massive spin 1 particles. These structures are required for writing the middle three-point amplitude. The three sub-amplitudes are then given by 
	\begin{eqnarray}
		\mathcal{A}_{1}\left(1,2,P_{1}^{(I_{1}I_{2})}\right) & = & \frac{g_{1}}{m_{1}}\langle 1P_{1}^{(I_{1}}\rangle\langle 2P_{1}^{I_{2})}\rangle[12]~,\nonumber\\
		\mathcal{A}_{2}\left(5,P_{1(I_{1}I_{2})},P_{2(J_{1}J_{2})}\right) & = & g_{21}[P_{1(I_{1}}5][P_{1 I_{2})}5]\langle 5P_{2(J_{1}}\rangle\langle5P_{2J_{2})}\rangle \nonumber\\
		&& + g_{22}[P_{1(I_{1}}5]\langle P_{1I_{2})}5\rangle[5P_{2(J_{1}}]\langle5P_{2J_{2})}\rangle\nonumber\\
		&& + g_{23}\langle P_{1(I_{1}}5\rangle\langle P_{1I_{2})}5\rangle[5P_{2(J_{1}}][5P_{2J_{2})}]~,\nonumber\\
		\mathcal{A}_{3}\left(3,4,P_{2}^{(J_{1}J_{2})}\right) & = & \frac{g_{3}}{m_{2}}\langle P_{2}^{(J_{1}}3\rangle\langle P_{2}^{J_{2})}4\rangle[34]~.
	\end{eqnarray}
	The residue can then be expressed as
	\begin{eqnarray}
		\mathcal{A}_{1}\mathcal{A}_{2}\mathcal{A}_{3} &=& \frac{g_{1}g_{3}}{m_{1}m_{2}}\biggl\{\frac{4g_{21}}{m_{2}^{2}}\langle12\rangle[34]\langle1|2|1]\langle3|4|3]^{2}[15][25]\langle35\rangle\langle45\rangle\nonumber\\
		&& \hspace{0.5cm} + \frac{g_{22}}{m_{1}m_{2}}\langle1|2|1]^{2}\langle3|4|3]^{2}\langle5|\left(p_{1}-p_{2}\right)|5]\langle5|\left(p_{4}-p_{3}\right)|5]\nonumber\\
		&&\hspace{0.8cm} + \frac{4g_{23}}{m_{1}^{2}}[12]\langle34\rangle\langle1|2|1]^{2}\langle3|4|3]\langle15\rangle\langle25\rangle[35][45]\biggr\}~.
	\end{eqnarray}
	Using the frame transformations, we find 
	\begin{eqnarray}
		\langle15\rangle = i\sqrt{s_{12}-s_{34}}e^{\frac{i}{2}\left(\phi-\psi\right)}\sin\frac{\theta_{12}}{2} = [15]^{\ast}~, &\quad& \langle25\rangle= \sqrt{s_{12}-s_{34}}e^{\frac{i}{2}\left(\phi+\psi\right)}\cos\frac{\theta_{12}}{2} = [25]^{\ast}~, \nonumber\\
		\langle35\rangle = \sqrt{s_{12}-s_{34}}e^{-\frac{i}{2}\left(\phi'+\psi'\right)}\sin\frac{\theta_{34}}{2}[35]^{\ast}~, &\quad& \langle45\rangle = -i\sqrt{s_{12}-s_{34}}e^{-i\left(\phi'-\psi'\right)}\cos\frac{\theta_{34}}{2}= [45]^{\ast}~.\nonumber
	\end{eqnarray}
	As explained in \ref{app:5ptparameter}, we can set $\psi =0= \psi'$ and $\omega=\phi+\phi'$. Using this fact, we can see
	\begin{eqnarray}\label{uneq-11}
		\mathcal{A}_{1}\mathcal{A}_{2}\mathcal{A}_{3} &=& \frac{g_{1}g_{3}}{m_{1}m_{2}}s_{12}s_{34}\left(s_{12}-s_{34}\right)^{2}\biggl\{\sqrt{s_{12}s_{34}}\sin\theta_{12}\sin\theta_{34}\left(g_{21}\frac{s_{34}}{m_{2}^{2}}e^{-i\omega}+g_{23}\frac{s_{12}}{m_{1}^{2}}e^{i\omega}\right)\nonumber\\
		&& \hspace{2cm} - g_{22}\frac{s_{12}s_{34}}{m_{1}m_{2}}\sin\theta_{12}\sin\theta_{34}\biggr\}~.
	\end{eqnarray}
	The above result reproduces the linear combination of the basis elements for spin 1 exchanges in both the channels,
	\begin{equation}
		a_{11-1}P_{1}^{1}\left(x\right)P_{1}^{1}\left(y\right)e^{-i\omega}+a_{110}P_{1}^{0}\left(x\right)P_{1}^{0}\left(y\right)+ a_{111}P_{1}^{1}\left(x\right)P_{1}^{1}\left(y\right)e^{i\omega}~.
	\end{equation}
	The terms inside the parentheses in \eqref{uneq-11} has been written in a suggestive manner - on the poles, $s_{12}=m_{1}^{2}$ and $s_{34}=m_{2}^{2}$, if the coupling constants are same, \textit{i.e.} $g_{21}=g_{23}$, then we can take their linear combination as a basis. In such cases\footnote{For later reference, we remark that whenever we will be using $\cos\left(n\omega\right)$ as the basis elements, the partial-wave coefficients will be denoted by $\tilde{a}^{(s_{12}, s_{34})}_{jln}$.}, instead of the basis spanned by $e^{in\omega}$ we can work with $\cos\left(n\omega\right)$, where $n\in \{0,1,\ldots\min\{j,l\}\}$. It can be checked that individual basis elements containing $e^{i\omega}$ gives rise to non-local terms in the kinematic variables,
	\begin{eqnarray}
		P_{1}^{1}\left(x\right)P_{1}^{1}\left(y\right)e^{i\omega} & \rightarrow & -\frac{2}{\sqrt{m_1 m_2} \left(m_1-m_2\right){}^2 }  \biggl\{\left(m_1-m_2\right) \left(\left(m_1-m_2\right) s_{23}+m_2
		s_{45}\right)\nonumber\\
		&&+\left(m_1-m_2\right) \biggl\{\left(\left(m_1-m_2\right) s_{23}+m_2
			s_{45}\right){}^2+s_{51}^2 \left(m_1-s_{45}\right){}^2\nonumber\\
			&&+2 s_{51} \left(-m_2
			s_{45}^2+s_{45} \left(\left(m_1+m_2\right) s_{23}+m_1 m_2\right)+m_1
			\left(m_2-m_1\right) s_{23}\right)\biggl\}^{\frac{1}{2}}\nonumber\\
			&& +s_{51} \left(\left(m_1+m_2\right) s_{45}+m_1
		\left(m_2-m_1\right)\right)\biggr\}~, \nonumber\\
			P_{1}^{1}\left(x\right)P_{1}^{1}\left(y\right)e^{-i\omega} & \rightarrow &
	8 \sqrt{m_1 m_2}  s_{45} s_{51} \left(m_1-m_2-s_{45}\right)
		\left(m_1-m_2+s_{51}\right)\nonumber\\
		&&  \biggl\{\left(m_1-m_2\right){}^2 \biggl\{\left(m_1-m_2\right)
			\left(\left(m_1-m_2\right) s_{23}+m_2 s_{45}\right)\nonumber\\
			&&+\left(m_1-m_2\right)
			\biggl\{\left(\left(m_1-m_2\right) s_{23}+m_2 s_{45}\right){}^2 +s_{51}^2
			\left(m_1-s_{45}\right){}^2 \nonumber\\
			&& +2 s_{51} \left(-m_2 s_{45}^2+s_{45}
			\left(\left(m_1+m_2\right) s_{23}+m_1 m_2\right)+m_1 \left(m_2-m_1\right)
			s_{23}\right)\biggr\}^{\frac{1}{2}} \nonumber\\
			&& +s_{51} \left(\left(m_1+m_2\right) s_{45}+m_1
			\left(m_2-m_1\right)\right)\biggr\}\biggr\}^{-1}~.
	\end{eqnarray}
	Only after adding both the terms above, we get local expressions of the form 
	\begin{equation}
		\frac{4 s_{51} \left(m_1 \left(m_1-m_2\right)-\left(m_1+m_2\right) s_{45}\right)-4
			\left(m_1-m_2\right) \left(\left(m_1-m_2\right) s_{23}+m_2
			s_{45}\right)}{\sqrt{m_1m_2} \left(m_1-m_2\right){}^2 }~.
	\end{equation}
	Note that \eqref{uneq-11} also implies that if the masses of the exchanged particles, having same spin are interchanged, then the residue remains unchanged.

	\paragraph{$j=1$ and $l=2$ :} Residue in this case is obtained by gluing the following sub-amplitudes,
	\begin{eqnarray}
		\mathcal{A}_{1}\left(1,2,P_{1}^{(I_{1}I_{2})}\right) & = & \frac{g_{1}}{m_{1}}\langle 1P_{1}^{(I_{1}}\rangle\langle2P_{1}^{I_{2})}\rangle[12]~,\nonumber\\
		\mathcal{A}_{2}\left(5,P_{1(I_{1}I_{2})},P_{2(J_{1}J_{2}J_{3}J_{4})}\right) & = & g_{21}\langle P_{1(I_{1}}5\rangle\langle P_{I_{2})}5\rangle\langle5P_{2(J_{1}}\rangle[5P_{2J_{2}}][5P_{2J_{3}}][5P_{2J_{4})}]\nonumber\\
		&& + g_{22}\langle P_{1(I_{1}}5\rangle[P_{1I_{2})}5]\langle 5P_{2(J_{1}}\rangle\langle5P_{2J_{2}}\rangle[5P_{2J_{3}}][5P_{2J_{4})}]\nonumber\\
		&& + g_{23} [P_{1(I_{1}}5][P_{1I_{2})}5]\langle5P_{2(J_{1}}\rangle\langle5P_{2J_{2}}\rangle\langle5P_{2J_{3}}\rangle[5P_{2J_{4})}]~, \nonumber\\
		\mathcal{A}_{3}\left(3,4,P_{2}^{(J_{1}J_{2}J_{3}J_{4})}\right) & = & \frac{g_{3}}{m_{3}^{3}}\langle P_{2}^{(J_{1}}3\rangle\langle P_{2}^{J_{2}}3\rangle\langle P_{2}^{J_{3}}4\rangle\langle P_{4}^{J_{4})}4\rangle[34]^{2}~.
	\end{eqnarray}
	The residue for exchanges of spin 1 state in $s_{12}$ and spin 2 state in $s_{34}$ channels is then 
	\begin{eqnarray}
		\mathcal{A}_{1}\mathcal{A}_{2}\mathcal{A}_{3} & = & \frac{4g_{1}g_{3}}{m_{1}m_{2}^{4}}\langle1|2|1]\langle3|4|3]^{3}\biggl\{\frac{g_{21}}{m_{1}^{2}}\langle1|2|1][12]\langle34\rangle\langle15\rangle\langle25\rangle[35][45]\langle5|\left(p_{3}-p_{4}\right)|5]\nonumber\\
		&& -\frac{g_{22}}{m_{1}m_{2}}\langle1|2|1]\langle3|4|3]\langle5|\left(p_{1}-p_{2}\right)|5]\left(\langle5|\left(p_{3}-p_{4}\right)|5]^{2}-2\langle5|3|5]\langle5|4|5]\right)\nonumber\\
		&& +\frac{g_{23}}{m_{2}^{2}}\langle3|4|3]\langle12\rangle[34][15][25]\langle35\rangle\langle45\rangle\langle5|\left(p_{3}-p_{4}\right)|5]\biggr\} \nonumber\\
		& = & \frac{g_{1}g_{3}}{m_{1}m_{2}^{4}}s_{12}s_{34}^{3}\left(s_{12}-s_{34}\right)^{3}\biggl\{\sqrt{s_{12}s_{34}}\sin\theta_{12}\sin\theta_{34}\cos\theta_{34}\left(g_{21}\frac{s_{12}}{m_{1}^{2}}e^{i\omega}+g_{23}\frac{s_{34}}{m_{2}^{2}}e^{-i\omega}\right)\nonumber\\
		&& \hspace{1cm} -2g_{22}\frac{s_{12}s_{34}}{m_{1}m_{2}}\cos\theta_{12}\left(-1+3\cos^{2}\theta_{34}\right)\biggr\}~.
	\end{eqnarray}
	The last equality can be expressed as the following linear combination
	\begin{equation}
		a_{12-1}P_{1}^{1}\left(x\right)P_{2}^{1}\left(y\right)e^{-i\omega}+a_{120}P_{1}^{0}\left(x\right)P_{2}^{0}\left(y\right)+ a_{121}P_{1}^{1}\left(x\right)P_{2}^{1}\left(y\right)e^{i\omega}~.
	\end{equation}
	\newline
	Residue of the five-point Veneziano amplitude, \eqref{Ven-5pt} with $m=0$, at the poles $s_{12}=1$ and $s_{34}=2$ is $r_{12}= \frac{1}{2} \left(s_{45}+1\right) \left(2 s_{23}+s_{45} \left(s_{51}-1\right)+2\right)$. Written in terms of the angular variables, this takes the form $-\frac{1}{16} \left(y+1\right) \left(4 \sqrt{2-2 x^2} \sqrt{1-y^2} \cos \omega +x \left(5y-1\right)-y-3\right)$. From this expression we can infer that spectrum of the exchanged states in that mass-level contains $j=0,1$ and $l=0,1,2$. 
	\begin{table}[h!]
		\centering
		\begin{tabular}{|c ccccccc|}
			\hline
			$(0,0,0)$ & $(0,1,0)$ & $(0,2,0)$ & $(1,0,0)$ & $(1,1,\pm 1)$ & $(1,1,0)$ & $(1,2,\pm1)$ & $(1,2,0)$\\
			\hline
			$\frac{5}{24}$ & $\frac{1}{4}$ & $\frac{1}{24}$ & $-\frac{1}{24}$ &$-\frac{1}{4\sqrt{2}}$ & -$\frac{1}{4}$ & $-\frac{1}{12\sqrt{2}}$ & $-\frac{5}{24}$\\
			\hline
		\end{tabular}
		\caption{Partial-wave coefficients $a_{jln}^{(1,2)}$ for  residue of massless five-point  Veneziano amplitude at  $s_{12}=1$,  $s_{34}=2$.}
	\end{table}
	\newline
	From the symmetry argument of \eqref{cs-5pt}, residue at poles $s_{12}=2$ and $s_{34}=1$ can be obtained by interchanging $s_{45}$ and $s_{51}$ in $r_{12}$. In this case spins of the exchanged states are $j=0,1,2$ and $l=0,1$. Partial-wave coefficients, $a_{jln}^{(1,2)}$ and $a_{jln}^{(2,1)}1$ can be seen to be related to each other in exactly the same way as $m_{1}$ and $m_{2}$ exchanges in the calculation of residues demonstrated above. 
	\begin{table}[h!]
		\centering
		\begin{tabular}{|c ccccccc|}
			\hline
			$(0,0,0)$ & $(0,1,0)$ & $(1,0,0)$ & $(1,1,\pm 1)$ & $(1,1,0)$ & $(2,0,0)$& $(2,1,\pm 1)$ & $(2,1,0)$ \\
			\hline
			$\frac{5}{24}$ & $-\frac{1}{24}$ & $\frac{1}{4}$ & $-\frac{1}{4\sqrt{2}}$ & $-\frac{1}{4}$ & $\frac{1}{24}$ & $-\frac{1}{12\sqrt{2}}$ & $-\frac{5}{24}$\\
			\hline
		\end{tabular}
		\caption{Partial-wave coefficients $a_{jln}^{(2,1)}$ for residue of massless five-point Veneziano amplitude at  $s_{12}=2$, $s_{34}=1$.}
	\end{table}
	
	\subsection{Equal mass exchanges}
	Here we consider masses of the exchanged states  in both the factorization channels, $s_{12}=s_{34}$ are same. Three-point kinematic forces $P_{1}\cdot p_{5} = 0 = P_{2}\cdot p_{5}$. In this case $E$ and $E'$ in \eqref{transf-sol} vanish, implying $p_{5}$ scales to zero. This is like a soft limit for the fifth particle's momentum.  
    
    Three-point amplitude with one massless and two massive legs of equal masses has the following basis structure\footnote{The amplitude is obtained by contracting $\lambda_{P_{1(I_{1}}}^{\alpha_{1}}\cdots \lambda_{P_{1}I_{2j})}^{\alpha_{2j}} \lambda_{P_{2}(J_{1}}^{\beta_{1}}\cdots \lambda_{P_{2}J_{2l)}}^{\beta_{2l}}$. Further details can be found in \cite{Arkani-Hamed:2017jhn}.}
	\begin{equation}
		\sum_{|j-l|}^{j+l}g_{i}x^{i}\left(\lambda_{5}^{2i}\varepsilon^{j+l-i}\right)_{\{\alpha_{1}\cdots\alpha_{2j}\},\{\beta_{1}\cdots\beta_{2l}\}}~.
	\end{equation}
    Here $x$ is the proportionality constant for $\frac{P_{1\alpha\dot{\alpha}}}{m}\tilde{\lambda}_{5}^{\dot{\alpha}}\propto \lambda_{5\alpha}$ and $\varepsilon_{\alpha\beta}$ is the quantity for $\mathbb{SL}\left(2,\mathbb{C}\right)$ contraction. 
	Since in the soft limit $\lambda_{5}\rightarrow 0$, any term proportional to $\lambda_{5}$ vanishes. Therefore the only term that survives is $i=0$ for $j=l$. For $j\ne l$ the three-point amplitude vanishes. This is consistent with the five-point massless Veneziano amplitude, where partial wave coefficients vanish for $j\ne l$ when $s_{12}=s_{34}$.
	
	\paragraph{Spin 1 exchange:} 
	The relevant three-point sub-amplitudes are
	\begin{eqnarray}
		\mathcal{A}_{1} & = & \frac{g}{m}\langle1P_{1}^{(I_{1}}\rangle\langle2P_{1}^{I_{2})}\rangle[12]~,\nonumber\\
		\mathcal{A}_{2}& = & g_{0}\langle P_{1(I_{1}}P_{2(J_{1}}\rangle\langle P_{1 I_{1})}P_{2 J_{2})}\rangle~,\nonumber\\
		\mathcal{A}_{3} & = & \frac{g}{m}\langle P_{2}^{(J_{1}}3\rangle\langle P_{2}^{J_{2})}4\rangle[34]~.
	\end{eqnarray}
    We denote the masses of the exchanged particles as $m$ and this is not to be confused with the external mass term appearing in \eqref{Ven-5pt}. 
	Gluing the three amplitudes, we get
	\begin{eqnarray}
		\mathcal{A}_{1}\mathcal{A}_{2}\mathcal{A}_{3} & = & \frac{g^{2} g_{0}}{m^{6}}[12][34]\langle1|2|1]^{2}\langle3|4|3]^{2} \biggl\{\langle13\rangle\langle24\rangle + \langle14\rangle\langle23\rangle\biggr\}\nonumber\\
		& = & -\frac{g^{2}g_{0}}{m^{6}}s^{6}\biggl\{\cos\theta_{12}\;\cos\theta_{34}+\cos\omega\; \sin\theta_{12}\;\sin\theta_{34}\biggr\}~.
	\end{eqnarray}
	We have denoted $s_{12}=s_{34}=s$. Interestingly, we observe that the terms inside the curly brackets come as a particular linear combination of  basis elements with specific coefficients, 
	\begin{equation}\label{equal-11}
		P_{1}^{0}\left(x\right)P_{1}^{0}\left(y\right) + P_{1}^{1}\left(x\right)P_{1}^{1}\left(y\right)\cos\left(\omega\right)~.
	\end{equation}
	As we show below, this feature holds for spin 2 case also. 
	
	\paragraph{Spin 2 exchange:}
	The relevant three-point amplitudes are the following,
	\begin{eqnarray}
		\mathcal{A}_{1}\left(1,2,P_{1}^{(I_{1}I_{2}I_{3}I_{4})}\right) & = & \frac{g}{m^{3}}\langle1P_{1}^{(I_{1}}\rangle\langle1P_{1}^{I_{2}}\rangle\langle2P_{1}^{I_{3}}\rangle\langle2P_{1}^{I_{4})}\rangle[12]^{2}~,\nonumber\\
		\mathcal{A}_{2}\left(5,P_{1\left(I_{1}I_{2}I_{3}I_{4}\right)}, P_{2\left(J_{1}J_{2}J_{3}J_{4}\right)}\right) & = & g_{0}\langle P_{1(I_{1}}P_{2(J_{1}}\rangle\langle P_{1I_{2}}P_{2J_{2}}\rangle\langle P_{1I_{3}}P_{2J_{3}}\rangle\langle P_{1I_{4})}P_{2J_{4})}\rangle~, \nonumber\\
		\mathcal{A}_{3}\left(3,4,P_{2(J_{1}J_{2}J_{3}J_{4})}\right) & = & \frac{g}{m^{3}}\langle P_{2(J_{1}}3\rangle\langle P_{2J_{2}}3\rangle\langle P_{2J_{3}}4\rangle\langle P_{2J_{4})}4\rangle[34]^{2}~.
	\end{eqnarray}
	The residue can be expressed as 
	\begin{eqnarray}
		\mathcal{A}_{1}\mathcal{A}_{2}\mathcal{A}_{3} & = & \frac{g^{2}g_{0}}{m^{14}}[12]^{2}[34]^{2}\langle1|2|1]^{4}\langle3|4|3]^{4}\biggl\{\langle13\rangle^{2}\langle24\rangle^{2}+\langle14\rangle^{2}\langle23\rangle^{2} + 4\langle13\rangle\langle14\rangle\langle23\rangle\langle24\rangle\biggr\} \nonumber\\
		& = & \frac{g^{2}g_{0}}{m^{14}}s^{12} \times \frac{1}{16}  \ \biggl\{2 \sin ^2\left(\theta _{12}\right) \left(6 \sin ^2\left(\theta _{34}\right) \cos
		(2 \omega )-1\right)+12 \sin \left(2 \theta _{12}\right) \sin \left(2 \theta _{34}\right) \cos (\omega )\nonumber\\
		&& \hspace{1cm} +8
		\cos ^2\left(\theta _{12}\right) \cos ^2\left(\theta _{34}\right)+\left(7 \cos \left(2 \theta
		_{12}\right)+1\right) \cos \left(2 \theta _{34}\right)\biggr\}~.
	\end{eqnarray}
	It can be checked that $1/16$ times the entire expression within the curly brackets can be expressed as 
	\begin{equation}\label{equal-22}
		P_{2}^{0}\left(x\right)P_{2}^{0}\left(y\right) + \frac{1}{3}P_{2}^{1}\left(x\right)P_{2}^{1}\left(y\right)\cos\left(\omega\right) + \frac{1}{12}P_{2}^{2}\left(x\right)P_{2}^{2}\left(y\right)\cos\left(2\omega\right)~.
	\end{equation}
	\newline
	In case of Veneziano amplitude \eqref{Ven-5pt} with massless external states, residue at $s_{12}=1$ and $s_{34}=1$ is given by $r_{11} = 1+s_{23} +s_{45}s_{51}$, which written in terms of the angular parameters takes the form $\frac{1}{2}\left(1-xy+\sqrt{1-x^{2}}\sqrt{1-y^{2}}\right)$. It is convenient to resort to the basis elements spanned by $P_{j}^{n}\left(x\right)P_{l}^{n}\left(y\right)\cos\left(n\omega\right)$. The non-vanishing partial-wave coefficients in this case are given in Table \ref{Tab:eqm-1}.
	\begin{table}[h!]
		\centering
		\begin{tabular}{|ccc|}
			\hline
			$\left(0,0,0\right)$ & $\left(1,1,0\right)$ & $\left(1,1,1\right)$ \\
			\hline
			$\frac{1}{2}$ & $-\frac{1}{2}$ & $-\frac{1}{2}$\\
			\hline
		\end{tabular}
		\caption{Non-zero partial-wave coefficients $\tilde{a}^{(1,1)}_{jln}$ for the residue of massless five-point Veneziano amplitude at $s_{12}=s_{34}=1$.}
        \label{Tab:eqm-1}
	\end{table}
	\newline
	Residue at the second mass-level, $s_{12}=s_{34}=2$ is $r_{22}= \frac{1}{4} \bigl\{2 s_{23}^2+\left(4 s_{45} s_{51}+6\right) s_{23}+s_{45} \left(s_{45}
	\left(s_{51}-1\right)-s_{51}+5\right) s_{51}+4\bigr\}$, which has the form $\frac{1}{2} \bigl\{\left(x^2-1\right) \left(y^2-1\right) \cos ^2(\omega )+\sqrt{1-x^2}
	\sqrt{1-y^2} (2 x y-1) \cos (\omega )+x y (x y-1)\bigr\}$ in terms of the angle parameters. The non-vanishing coefficients for this case are given in Table \ref{Tab:22}. 
	\begin{table}[h!]
		\centering
		\begin{tabular}{|cccccc|}
			\hline
			$\left(0,0,0\right)$ & $\left(1,1,0\right)$ & $\left(1,1,1\right)$ & $\left(2,2,0\right)$ & $\left(2,2,1\right)$ & $\left(2,2,2\right)$ \\
			\hline
			$\frac{1}{6}$ & $-\frac{1}{2}$ & $-\frac{1}{2}$ & $\frac{1}{3}$ & $\frac{1}{9}$ & $\frac{1}{36}$\\
			\hline
		\end{tabular}
		\caption{Non-zero partial-wave coefficients $\tilde{a}^{(2,2)}_{jln}$ for the residue of massless five-point Veneziano amplitude at $s_{12}=s_{34}=2$.}
        \label{Tab:22}
	\end{table}
    \newline
	Note that in the tables above, the relative coefficients of spin 1 exchange at both the mass-levels appear in the same ratio as in \eqref{equal-11} and \eqref{equal-22}, \textit{i.e.} $\tilde{a}_{110}:\tilde{a}_{111} = 1:1$. Also from Table \ref{Tab:22} we see $\tilde{a}_{220}:\tilde{a}_{221}:\tilde{a}_{222}= 1:\frac{1}{3}:\frac{1}{12}$ which corroborates \eqref{equal-22}.
	
	\section{Constraints on five-point coefficients}
    \label{Sec:5pt-Constr}
    In this section we derive relations between the partial-wave coefficients for residues of five-point and four-point amplitudes. These relations can in turn be used to fix some of the five-point coefficients using  the four-point ones.  
	\subsection{Lowest mass-level}
	We have already checked that the single-channel residue of \eqref{Ven-5pt} is equal to the four-point Veneziano amplitude,
	\begin{equation}
		\text{Res}_{s_{34}=m^{2}}\mathcal{M}_{5}^{\text{Ven}} = \mathcal{M}_{4}^{\text{Ven}}\left(s_{12}, s_{51}\right)~, \qquad \text{Res}_{s_{12}=m^{2}}\mathcal{M}_{5}^{\text{Ven}} = \mathcal{M}_{4}^{\text{Ven}}\left(s_{34}, s_{45}\right)~.
	\end{equation}
	Solving \eqref{kinem-param}, we can see
	\begin{equation}
		x=1+\frac{2s_{51}}{s_{12}-4m^{2}}, \quad \text{when}\; s_{34}=m^{2}~; \qquad y=1+\frac{2s_{45}}{s_{34}-4m^{2}}, \quad \text{when}\; s_{12}=m^{2}~.
	\end{equation}
	On the lowest mass-level, only spin 0 state propagates. Therefore, residue of the five-point amplitude on the poles $s_{12}=m^{2}+k$ with $k$ being a positive integer and $s_{34}=m^{2}$ has the expansion
	\begin{equation}
		\text{Res}_{s_{12}=m^{2}+k, s_{34}=m^{2}}\mathcal{M}_{5}^{\text{Ven}} = \sum_{j}a^{(k,0)}_{j00}\mathcal{G}^{\left(\frac{d-3}{2}\right)}_{j}\left(x\right)~.
	\end{equation}
	 On the other hand, residue of the four-point amplitude on  the poles of $s_{12}$ have the expansions in Gegenbauer polynomials with the known four-point partial wave coefficients,
	\begin{equation}
		\text{Res}_{s_{12}=m^{2}+k}\mathcal{M}_{4}^{\text{Ven}}\left(s_{12}, s_{51}\right) = \sum_{j} c^{(k)}_{j}\mathcal{G}^{\left(\frac{d-3}{2}\right)}_{j}\left(\tilde{x}\right), \qquad \tilde{x}=1+\frac{2s_{51}}{k-3m^{2}}~.
	\end{equation} 
	Comparing the two expansions we get the constraints,
	\begin{equation}
		a^{(k,0)}_{j00}=a^{(0,k)}_{0j0}=c^{(k)}_{j}~.
	\end{equation}

	\subsection{Splitting relations}
    \label{Sec:splitting}
	 A remarkable observation about factorization of amplitude on zero loci of some kinematic variables for certain class of theories, such as $\text{tr}\left(\phi^{3}\right)$ and Yang-Mills theory, were made in \cite{Arkani-Hamed:2023swr}. 
	 For the five-point $\phi^{3}$ amplitude given in \eqref{phi3} it can be immediately seen that
	 \begin{eqnarray}
	 	\mathcal{M}_{5}^{\phi^{3}}|_{s_{45}=s_{12}+s_{23}-m^{2}} & =& \left(\frac{1}{s_{12}-m^{2}}+\frac{1}{s_{51}-m^{2}}\right)\times\left(\frac{1}{s_{23}-m^{2}}+\frac{1}{s_{34}-m^{2}}\right)~, \nonumber\\
	 	\mathcal{M}_{5}^{\phi^{3}}|_{s_{23}=s_{45}+s_{51}-m^{2}} & =& \left(\frac{1}{s_{12}-m^{2}}+\frac{1}{s_{51}-m^{2}}\right)\times\left(\frac{1}{s_{34}-m^{2}}+\frac{1}{s_{45}-m^{2}}\right)~.
	 \end{eqnarray} 
	 This factorization property of the five-point amplitude to four-point amplitudes on special values of certain kinematic variables is called splitting. Considering masses of all the external states to be $m$, then conservation of momenta implies
	 \begin{eqnarray}
	 	s_{13}+s_{12}+s_{23}-s_{45} & = & 3m^{2}~, \nonumber\\
	 	s_{14}+s_{45}+s_{51}-s_{23} & = & 3m^{2}~.
	 \end{eqnarray}
	Using the above equations, it can be checked that splitting relations translate to the following factorizations of the five-point Veneziano amplitude,
	 \begin{eqnarray}\label{hidden0}
	 	\text{On}\; s_{13}=2m^{2}, \quad \mathcal{M}_{5}^{\text{Ven}}\left(s_{45}=s_{12}+s_{23}-m^{2}\right) &=& \mathcal{M}_{4}^{\text{Ven}}\left(s_{12},s_{51}\right)\times\mathcal{M}_{4}^{\text{Ven}}\left(s_{23},s_{34}\right)~, \nonumber\\
	 	\text{On}\; s_{14}=2m^{2}, \quad \mathcal{M}_{5}^{\text{Ven}}\left(s_{23}=s_{45}+s_{51}-m^{2}\right) &=&\mathcal{M}_{4}^{\text{Ven}}\left(s_{12}, s_{51}\right)\times\mathcal{M}_{4}^{\text{Ven}}\left(s_{34}, s_{45}\right)~.\nonumber\\
	 \end{eqnarray}
	 At the level of residues, \eqref{hidden0} provide set of relations between the five-point and four-point coefficients. This principle can be used as a consistency check for the factorization of an arbitrary planar-ordered five-point amplitude. Furthermore, on the  loci $s_{13}=s_{14}=2m^{2}$ kinematic variables satisfy $s_{12}+s_{51}=2m^{2}$. This condition leads to the existence of hidden zeroes for the class of amplitudes for which $\mathcal{M}_{4}\left(s, 2m^{2}-s\right)=0$; the five-point amplitude vanishes on this kinematic constraint. In our analysis, solutions to the partial-wave coefficients obtained below, satisfy the conditions of hidden zeroes - the five-point residues vanish when the kinematic constraint is imposed. 

     In the following analysis, we will consider a putative five-point amplitude with integer-spaced spectrum which satisfy the splitting relations of \eqref{hidden0}. External states are taken to be identical massive scalars. In addition, similar to the Veneziano amplitude, we assume that at a mass-level, say $k$, spins of the exchanged particles range from $0$ to $k$ with unit interval. The analysis for the rest of this section holds in four dimensions only. We will stick to the basis containing $\cos\left(n\omega\right)$ and use $a^{(k_{1}, k_{2})}_{jln}$ to denote the coefficients for the residue at mass-level $s_{12}=m^{2} + k_{1}$ and $s_{34}=m^{2}+k_{2}$.

\subsubsection{Equal mass exchanges}
	 Here we consider same mass-level for $s_{12}$ and $s_{34}$. If we invert the relations in \eqref{kinem-param} with $s_{12}=s_{34}=s$,  we find
	 \begin{eqnarray}
	     x & = & \sqrt{\frac{s}{m^{2}\left(s-4m^{2}\right)\left(m^{2}-4s\right)}}\left(-3m^{2}+2s_{51}\right)~, \nonumber\\
         y &= & \sqrt{\frac{s}{m^{2}\left(s-4m^{2}\right)\left(m^{2}-4s\right)}}\left(-3m^{2}+2s_{45}\right)~.
	 \end{eqnarray}
	 In the limit $m\rightarrow 0$, $x\approx \frac{s_{51}}{\sqrt{-m^{2}s}}$ and $y\approx \frac{s_{45}}{\sqrt{-m^{2}s}}$. It may appear that cosines of the angles are divergent. However, note that in the massless limit $p_{5}\rightarrow 0$, which implies that $s_{45}$ and $s_{51}$ also vanish. Therefore $x$ and $y$  remain finite but undetermined. In this limit $\cos\omega\rightarrow 1$. To handle the cases of equal mass exchanges, it is necessary to retain the mass term and then take the limit $m\rightarrow 0$ at the end of the calculation to get the result for massless external states. In doing so, we have to be careful about the basis elements because for generic $m$, residues at $s_{12}=s_{34}$ contain basis elements with $j\ne l$, which are absent in the massless case. 
	 
	 \paragraph{First mass-level:} 
	 This is the case with $s_{12}=s_{34}=m^{2}+1$. We consider an arbitrary five-point residue at this mass-level spanned by undetermined coefficients,
     \begin{equation}
         a^{(1,1)}_{000}~, \quad a^{(1,1)}_{010}~,\quad a^{(1,1)}_{100}~, \quad a^{(1,1)}_{110}~, \quad a^{(1,1)}_{111}~,
     \end{equation}
     and the four-point undetermined coefficients are $c^{(1)}_{0}$ and $c^{(1)}_{1}$. We will work out the constraints coming from splitting conditions for this case in detail. 
     
     On the locus $s_{13}=2m^{2}$ we have $s_{45}=s_{12}+s_{23}-m^{2}$ and on the locus $s_{14}=2m^{2}$ we have $s_{23}=s_{45}+s_{51}-m^{2}$. The resulting splitting constraints for the massive $(1,1)$ residue are algebraically lengthy; for completeness we collect the explicit locus equations and intermediate relations in Appendix~\ref{app:splitting-algebra} (see Eqs.~\eqref{constr1-11}, \eqref{constr2-11} and \eqref{rel1-22}--\eqref{rel6-22}).

\begin{equation}\label{cntr11-1}
	 	c^{(1)}_{1} = \frac{1-3m^{2}}{1+m^{2}}c^{(1)}_{0}~, \qquad a^{(1,1)}_{110} = \frac{-2+3m^{2}+9m^{4}}{\left(1+m^{2}\right)^{2}}\left(c^{(1)}_{0}\right)^{2}~.
	 \end{equation}
	 Then substituting the above relations in the above equations, we obtain
	 \begin{eqnarray}\label{cnstr11-2}
	 	a^{(1,1)}_{000} & = & \frac{2+m^{2}}{1+m^{2}}\left(c^{(1)}_{0}\right)^{2}~,\nonumber\\
	 	a^{(1,1)}_{010} & = & a^{(1,1)}_{100} = \frac{\sqrt{-m^{2}\left(4-9\left(m^{2}+m^{4}\right)\right)}}{\left(1+m^{2}\right)^{\frac{3}{2}}}\left(c^{(1)}_{0}\right)^{2}~.
	 \end{eqnarray}
    To verify \eqref{cntr11-1} and \eqref{cnstr11-2} we take Veneziano amplitude as an example. The residue for the five-point Veneziano amplitude with massive external states in this case is $m^4-m^2 \left(s_{45}+s_{51}+1\right)+s_{23}+s_{45}
	 s_{51}+1$. In terms of the angle parameters, it takes the form
	 \begin{eqnarray*}
	 	&&\frac{1}{4} \biggl\{m^4 (9 x y+1)+2 \left(3 m^2-1\right)
	 	\sqrt{1-x^2} \sqrt{1-y^2} \cos (\omega )+3 m^2 (x
	 	y+1)\nonumber\\
	 	&&+\sqrt{-m^2 \left(m^2+1\right)} \sqrt{4-9
	 		\left(m^4+m^2\right)} x+\sqrt{-m^2
	 		\left(m^2+1\right)} \sqrt{4-9 \left(m^4+m^2\right)}
	 	y-2 x y+2\biggr\}~.
	 \end{eqnarray*}
	 Non-vanishing partial-wave coefficients for the massive case are given in Table \ref{Tab:11m}. 
	 \begin{table}[h!]
	 	\centering
	 	\begin{tabular}{|c|c|}
	 		\hline
	 		$\left(0,0,0\right)$ & $\frac{1}{4}\left(2+3m^{2}+m^{4}\right)$\\
	 		&\\
	 		\begin{tabular}{c}
	 			$\left(0,1,0\right)$  \\
	 			$\left(1,0,0\right)$
	 		\end{tabular}
	 		& $\frac{1}{4}\sqrt{-m^{2}\left(1+m^{2}\right)\left(4-9m^{2}-9m^{4}\right)}$\\
	 		&\\
	 		$\left(1,1,0\right)$ & $-\frac{1}{4}\left(2-3m^{2}-9m^{2}\right)$ \\
	 		&\\
	 		$\left(1,1,1\right)$ & $-\frac{1}{2}\left(1-3m^{2}\right)$\\
	 		\hline
	 	\end{tabular}
	 	\caption{Non-zero partial-wave coefficients $\tilde{a}^{(1,1)}_{jln}$ for the residue of five-point Veneziano amplitude at $s_{12}=s_{34}=m^{2}+1$.}
        \label{Tab:11m}
	 \end{table}
	 \newline
	 In the massless limit $\tilde{a}^{(1,1)}_{010}$ and $\tilde{a}^{(1,1)}_{100}$ vanish. On the other hand the four-point residue of the Veneziano amplitude has the partial-wave coefficients 
	 \begin{equation}\label{pw4-1}
	 	c^{(1)}_{0} = \frac{1}{2}\left(1+m^{2}\right)~, \qquad c^{(1)}_{1}=\frac{1}{2}\left(1-3m^{2}\right)~. 
	 \end{equation}
     Therefore \eqref{pw4-1} and the data given in Table. \eqref{Tab:11m} satisfy \eqref{cntr11-1} and \eqref{cnstr11-2}, implying that Veneziano amplitude passes the test.
     
 \paragraph{Second mass-level}
	 For the five-point residue, there are fourteen possible basis elements with coefficients, 
	 \begin{eqnarray}
	 	a^{(2,2)}_{000}~, \quad a^{(2,2)}_{010}~, \quad a^{(2,2)}_{020}~, \quad a^{(2,2)}_{100}~, \quad a^{(2,2)}_{110}~, \quad a^{(2,2)}_{111}~, \quad a^{(2,2)}_{120}~, \nonumber\\
	 	a^{(2,2)}_{121}~, \quad a^{(2,2)}_{200}~, \quad a^{(2,2)}_{210}~, \quad a^{(2,2)}_{211}~, \quad a^{(2,2)}_{220}~, \quad a^{(2,2)}_{221}~, \quad a^{(2,2)}_{222}~.
	 \end{eqnarray}
	 The four-point residue has three coefficients,
	 \begin{equation}
	 	c^{(2)}_{0}~, \quad c^{(2)}_{1}~, \quad c^{(2)}_{2}~.
	 \end{equation}
	 Next step is to repeat the same analysis done for the previous case of $s_{12}=s_{34}=1$. Here we  choose to keep the external mass small  for ease of calculation. Constraints from the splitting conditions to leading order in $m$ lead to the following relations,
	 \begin{eqnarray}\label{split22}
	 	a^{(2,2)}_{110} & = &  a^{(2,2)}_{111} = -c^{(2)}_{1}\left(c^{(2)}_{0}+c^{(2)}_{1} + c^{(2)}_{2}\right)~,\nonumber\\
	 	a^{(2,2)}_{220} & = & 3a^{(2,2)}_{221} = 12 a^{(2,2)}_{222} = -a^{(2,2)}_{000} + \left(c^{(2)}_{0}+c^{(2)}_{2}\right)\left(c^{(2)}_{0}+c^{(2)}_{1}+c^{(2)}_{2}\right)~,
	 \end{eqnarray}
	 and $a^{(2,2)}_{010}$, $a^{(2,2)}_{020}$, $a^{(2,2)}_{100}$, $a^{(2,2)}_{120}$, $a^{(2,2)}_{121}$, $a^{(2,2)}_{200}$, $a^{(2,2)}_{210}$ and $a^{(2,2)}_{211}$ are $\mathcal{O}\left(m\right)$. As a consistency check, equations in \eqref{split22} can be seen to satisfy the data presented in Table \eqref{Tab:22} along with the   coefficients for four-point Veneziano amplitude with $m=0$, 
	 \begin{equation}
	 	c^{(2)}_{0} = \frac{1}{6}~, \quad c^{(2)}_{1} = \frac{1}{2}~, \quad c^{(2)}_{2} = \frac{1}{3}~.
	 \end{equation}

     \subsubsection{Unequal mass exchanges}
	 At mass-level $s_{12}=m^{2}+1$ and $s_{34}=m^{2}+2$, there are eight coefficients for a five-point residue,
	 \begin{eqnarray}
	 	a^{(1,2)}_{00}~, \quad a^{(1,2)}_{010}~, \quad a^{(1,2)}_{020}~, \quad a^{(1,2)}_{100}~, \nonumber\\
	 	a^{(1,2)}_{110}~, \quad a^{(1,2)}_{111}~, \quad a^{(1,2)}_{120}~, \quad a^{(1,2)}_{121}~,
	 \end{eqnarray}%
	 whereas the corresponding four-point coefficients are 
	 \begin{equation}
	 	c^{(1)}_{0}~, \quad c^{(1)}_{1}~, \quad c^{(2)}_{0}~, \quad c^{(2)}_{1}~, \quad c^{(2)}_{2}~.
	 \end{equation}
	From the splitting conditions, we get the constraint on four-point coefficients at the first mass-level: $c^{(1)}_{1} = \frac{1-3m^{2}}{1+m^{2}}c^{(1)}_{0}$. In this case all the eight five-point coefficients get fixed by the four-point ones.

\noindent\textit{Explicit solution.} The closed-form expressions for the eight coefficients $a^{(1,2)}_{jln}$ obtained from the splitting constraints are lengthy; we present them in Appendix~\ref{app:splitting-algebra} as Eq.~\eqref{split-12}.
	It can be verified that \eqref{split-12} corroborates the values for Veneziano amplitude presented in Table \ref{Tab:pw-12}.
	\begin{table}[h!]
		\centering
		\begin{tabular}{|c|c|}
			\hline
			$\left(0,0,0\right)$ & $\frac{\left(m^2+1\right) \left(6 m^6+15 m^4+9
				m^2+10\right)}{24 \left(m^2+2\right)}$\\
				$\left(0,1,0\right)$ & $\frac{\sqrt{2-3 m^2} \left(m^4+3 m^2+2\right)
					\sqrt{1-3 m^2 \left(m^2+2\right)}}{8 \sqrt{m^2+2}}$\\
					$\left(0,2,0\right)$ & $\frac{\left(m^2+1\right) \left(9 m^6+12 m^4-15
						m^2+2\right)}{24 \left(m^2+2\right)}$\\
						$\left(1,0,0\right)$ & $\frac{\sqrt{1-3 m^2} \sqrt{m^2+1} \left(6 m^4+9
							m^2-2\right) \sqrt{1-3 m^2 \left(m^2+2\right)}}{24
							\left(m^2+2\right)}$\\
							$\left(1,1,0\right)$ & $-\frac{1}{8} \sqrt{1-3 m^2} \sqrt{2-3 m^2}
							\sqrt{m^2+1} \sqrt{m^2+2} \left(3 m^2+1\right)$ \\
							$\left(1,1,1\right)$ & $-\frac{1}{4} \sqrt{1-3 m^2} \sqrt{2-3 m^2}
							\left(m^2+1\right)$\\
							$\left(1,2,0\right)$ & $\frac{\sqrt{1-3 m^2} \sqrt{m^2+1} \sqrt{1-3 m^2
									\left(m^2+2\right)} \left(9
								\left(m^4+m^2\right)-10\right)}{24
								\left(m^2+2\right)}$\\
								$\left(1,2,1\right)$ & $\frac{\sqrt{1-3 m^2} \left(3 m^2-2\right) \sqrt{1-3
										m^2 \left(m^2+2\right)}}{12 \sqrt{m^2+2}}$\\
									\hline
		\end{tabular}
		\caption{Partial-wave coefficients $\tilde{a}^{(1,2)}_{jln}$ for the residue of five-point Veneziano amplitude at $s_{12}=1$, $s_{34}=2$.}
        \label{Tab:pw-12}
	\end{table}
	Four-point coefficients for Veneziano amplitude at the second mass-level are
	\begin{eqnarray}
		c^{(2)}_{0} & = & \frac{1}{12} \left(6 m^4-3 m^2+2\right)~, \nonumber\\
		c^{(2)}_{1} & = & \frac{1}{4} \left(-3 m^4-m^2+2\right)~, \nonumber\\
		c^{(2)}_{2} & = & \frac{1}{12} \left(2-3 m^2\right)^2~,
	\end{eqnarray}
	and that at the first mass-level are given in \eqref{pw4-1}.

    \paragraph{Higher mass-levels :} 
    We have repeated the analysis for the first few mass-levels for the massless case $m=0$. At $s_{12}=1$ and $s_{34}=3$, there are eleven coefficients associated to the five-point residue.  At this mass-level all the eleven coefficients get fixed by the four-point ones. The solutions are given in \eqref{split-13}. Similar conclusion holds for the mass-level $s_{12}=1$ and $s_{34}=4$. 

    At the mass-level $s_{12}=2$ and $s_{34}=3$, we find that not all five-point coefficients can be determined given the four-point ones. There are twenty partial-wave coefficients for the five-point residue, of which eighteen can be fixed using the remaining two and the four-point coefficients. These solutions are given in \eqref{split-23}. We have chosen $a^{(2,3)}_{000}$ and $a^{(2,3)}_{110}$ as the free data. There is also a constraint satisfied by the four-point coefficients at this mass-level, 
    \begin{equation}
 		c^{(2)}_{2} = c^{(2)}_{1} - c^{(2)}_{0}~.
 	\end{equation}
    In the case of $s_{12}=s_{34}=2$, we have seen from \eqref{split22} that one of the five-point coefficients remains unfixed. Based on these observations, we infer that whenever both the channels contain intermediate states with spins equal to or greater than two, splitting relations can not constrain all the five-point coefficients in terms of the four-point ones. 

    \subsection{Hidden zeros}
	 A useful corollary of our analysis is that \emph{splitting implies
the hidden zero}: if the absorptive part of the amplitude satisfies the splitting identities on both hypersurfaces in
\eqref{hidden0}, then it necessarily vanishes on their intersection (for $m=0$)
$s_{13}=s_{14}=0$.  We find it particularly illuminating that this implication is
manifest in partial-wave space: once the five-point partial-wave coefficients are constrained to
obey the two splitting relations, the resulting residue automatically vanishes on the intersection. This is true even for the solutions with a nontrivial kernel, indicating a larger class of Veneziano-like amplitudes having this property.

\normalcolor
	\subsection{Interpreting the four-point consistency conditions}
\label{sec:c-consistency}

In our explicit five-point analysis for the $m=0$ case, the compatibility of imposing two independent splitting loci forces
simple relations among the four-point partial-wave coefficients $c^{(k)}_j$.
For the first two mass levels we found
\begin{equation}
c^{(1)}_0=c^{(1)}_1,
\qquad
c^{(2)}_0-c^{(2)}_1+c^{(2)}_2=0.
\label{eq:c-constraints-12}
\end{equation}
These relations admit a unified interpretation in terms of the \emph{four-point residue polynomial}
at level $k$.

\paragraph{Residues as polynomials in the scattering angle.}
For a planar-ordered massless four-point amplitude, the residue on the pole $s=k$ may be expanded as
\begin{equation}
R_k(z)\;:=\;\Res_{s=k}M_4(s,t)
=\sum_{j=0}^{k}c^{(k)}_{j}\,P_j(z),
\qquad
z:=\cos\theta=1+\frac{2t}{k}\,,
\label{eq:Rk-def}
\end{equation}
where $P_j$ are Legendre polynomials (using $d=4$).
Evaluating \eqref{eq:Rk-def} at $z=-1$ and using $P_j(-1)=(-1)^j$ gives the identity
\begin{equation}
R_k(-1)=\sum_{j=0}^{k}(-1)^j\,c^{(k)}_j.
\label{eq:Rkminus1}
\end{equation}
Thus the constraints \eqref{eq:c-constraints-12} are precisely the statements that
\begin{equation}
R_1(-1)=0,\qquad R_2(-1)=0,
\label{eq:R1R2zero}
\end{equation}
and more generally the pattern suggests the level-$k$ condition
\begin{equation}
R_k(-1)=0
\qquad\Longleftrightarrow\qquad
\sum_{j=0}^k(-1)^j\,c^{(k)}_j=0.
\label{eq:Rk-hiddenzero}
\end{equation}

\paragraph{Physical meaning: a four-point hidden zero at $u=0$.}
For massless $2\to2$ kinematics at fixed $s=k$, the backward-scattering point $z=-1$ corresponds to
\begin{equation}
z=-1\quad\Longleftrightarrow\quad t=-k
\quad\Longleftrightarrow\quad u:=-s-t=0.
\end{equation}
Therefore \eqref{eq:Rk-hiddenzero} states that the level-$k$ residue has a \emph{zero} when the
``missing'' channel invariant vanishes, $u=0$.  In this sense the five-point compatibility conditions
\eqref{eq:c-constraints-12} enforce a four-point hidden-zero property already at $k=1,2$.
This also clarifies the mechanism behind the five-point hidden zero observed in our residue analyses:
once two different splitting relations are imposed, their intersection forces one of the four-point
factors to sit at $u=0$, and the five-point residue vanishes automatically.

\paragraph{EFT translation for planar colored scalars.}
In a low-energy expansion for a fixed color ordering,
\begin{equation}
M_4(s,t)=\sum_{n,m\ge0}\alpha_{n,m}\,s^n t^m,
\label{eq:EFT-exp-again}
\end{equation}
a pole contribution at $s=k$ takes the form
\begin{equation}
M_4(s,t)\supset \frac{R_k(t)}{k-s}
=\frac{1}{k}\,R_k(t)\sum_{n=0}^\infty\Big(\frac{s}{k}\Big)^n.
\label{eq:pole-expand}
\end{equation}
Hence the residue polynomial $R_k(t)$ generates, order-by-order in $s^n$, linear relations among the
EFT coefficients $\alpha_{n,m}$ contributed by that pole.
For $k=1$, \eqref{eq:c-constraints-12} implies $R_1(z)\propto(1+z)$, equivalently $R_1(t)\propto(1+t)$,
which yields the tower (for the $s=1$ pole contribution)
\begin{equation}
\alpha_{n,0}=\alpha_{n,1}\qquad (n\ge0).
\label{eq:alpha-k1}
\end{equation}
For $k=2$, the condition $R_2(-1)=0$ implies that $R_2(t)$ has a root at $t=-2$ and can be written as
$R_2(t)=(t+2)(A+Bt)$, which in turn implies (again for each fixed $n$ in \eqref{eq:pole-expand})
\begin{equation}
\alpha_{n,1}-\frac{1}{2}\,\alpha_{n,0}=2\,\alpha_{n,2}\qquad (n\ge0)
\qquad\text{(from the $s=2$ pole contribution)}.
\label{eq:alpha-k2}
\end{equation}
Equations \eqref{eq:alpha-k1}--\eqref{eq:alpha-k2} provide a simple EFT interpretation of the first
two consistency conditions in \eqref{eq:c-constraints-12}; higher levels are expected to impose
analogous linear relations associated with the backward-scattering/hidden-zero property
\eqref{eq:Rk-hiddenzero}.

\section{Discussion}
\label{sec:discussion}
In this paper, we have presented a partial-wave basis for the residues of five-point tree-amplitudes. The partial-wave analysis contains information about the spins of the exchanged states in a scattering, using which we learn about the spectrum of the theory under consideration. Moreover, the partial-wave coefficients are related to the coupling constants of various three-point functions. Unlike the case of four-point identical scalar scattering, at five-point we find two types of cubic vertices required to calculate the residues. The basis presented in \eqref{pwexp-d} holds for identical external states. It is possible to construct a more general basis involving spinning external states by gluing allowed three-point structures, which are classified in \cite{Boels:2012ie, Boels:2012if, Chakraborty:2020rxf, Caron-Huot:2022jli}. We leave this for future work.  The basis presented here is particularly important for dual amplitudes, which can be expressed as sum of the residues over poles.  It will be interesting to generalize this partial-wave basis beyond tree-level. One possible way is to implement generalized unitarity \cite{Bern:2011qt} to construct loop amplitudes from the residues.  

We summarize the observations made in this paper and point to possible future explorations below. 

\begin{itemize}

\item \textbf{Rigidity from five-point splitting in partial-wave space:}
Our main conceptual takeaway is that five-point splitting/hidden-zero constraints become \emph{linear} relations among the five-point partial-wave coefficients $a^{(k_1,k_2)}_{jln}$ e.g. \eqref{split22}, while simultaneously inducing \emph{compatibility} conditions on the four-point data $c^{(k)}_j$ - \eqref{eq:c-constraints-12}.
In the lowest nontrivial examples, two independent splitting loci together with the expected spin truncation fix all five-point coefficients uniquely in terms of the four-point ones (e.g.\ $(1,2)$ and $(1,3)$), making the higher-point constraints strikingly transparent in partial-wave language \cite{Arkani-Hamed:2023swr,Berman:2025splittingregions}.

\item \textbf{Physical meaning of the four-point compatibility conditions:}
The relations we find among the four-point partial-wave data (e.g.\ $c^{(1)}_0=c^{(1)}_1$ and
$c^{(2)}_0-c^{(2)}_1+c^{(2)}_2=0$) are best viewed as \emph{compatibility} conditions required for a
given five-point residue to satisfy two independent splitting identities simultaneously.  In fact,
both constraints admit a unified interpretation: they are equivalent to the statement that the
corresponding four-point residue polynomial $R_k(z)=\sum_{j=0}^k c^{(k)}_j P_j(z)$ vanishes at
backward scattering, $R_k(-1)=0$ (equivalently, at $u=0$ for massless kinematics).  In an EFT
interpretation for planar-ordered colored scalars this translates into structured linear relations
among low-energy expansion coefficients sourced by each pole level, thereby providing an analytic
mechanism for how higher-point splitting/hidden-zero constraints can propagate back onto four-point
data and shrink the allowed region in Wilson-coefficient space \cite{Berman:2025splittingregions}. The present analysis applies to tree-level amplitudes which contain simple poles, it is worth extending it to general amplitudes with loops. For this we have to generalize the splitting conditions for generic discontinuities of the amplitudes such as branch cuts and also we need to formulate the five-point partial wave basis in the presence of loops. It will be interesting to understand if the splitting relations can be translated to some consistency conditions for having particular high energy behavior of the amplitudes.

\item \textbf{The first spin-2 obstruction and the emergence of a kernel:}
Beginning at the first mass level where spin-2 exchange is allowed (illustrated by the $(2,3)$ residue), the two basic five-point splitting loci no longer suffice to determine all $a^{(k_1,k_2)}_{jln}$: a genuine residual kernel remains, naturally supported in the new higher-$n$ harmonics (notably the $n=2$ sector).
This clarifies why the low-level uniqueness seen at $(1,2)$ and $(1,3)$ should not be expected to persist automatically once the angular/harmonic content grows.

\item \textbf{How the kernel might be eliminated - additional higher-point input:}
The persistence of a kernel at spin $\ge2$ suggests that complete rigidity requires information beyond the two standard five-point splitting loci.
Natural candidates include:
\begin{itemize}
\item \emph{additional vanishing/splitting loci} (``hidden-zero'' constraints beyond the two splits) \cite{Arkani-Hamed:2023swr};
\item \emph{higher-multiplicity constraints}, most notably six-point splitting relations, which can feed back into five-point data and potentially remove residual five-point ambiguities, in the spirit of higher-point bootstrap constraints \cite{Berman:2025splittingregions,apratim6};
\item \emph{mixed-multiplicity positivity constraints} (``multipositivity'') which provide an infinite family of inequalities simultaneously constraining residues and EFT coefficients at multiple multiplicities, and are known to be sharply saturated by open-string tree amplitudes \cite{Cheung:2025Multipositivity}.
\end{itemize}
On the other hand, it is entirely likely that this lack of rigidity will enable us to enlarge the space of string-like theories; a thought that should not be ignored!

\item \textbf{Unitarity and analyticity beyond four points: what partial waves do (and do not) solve:}
While our partial-wave basis provides a clean harmonic decomposition of five-point residues and a natural arena to phrase factorization constraints, imposing full multiparticle unitarity remains intrinsically nonlinear and couples different multiplicities through phase-space integrals at loop-level.
A promising direction is to reformulate multi-particle cuts and their positivity properties directly in the $(j,l,n)$ basis, potentially interfacing our framework with recent positivity-driven bootstrap developments \cite{Cheung:2025Multipositivity,Berman:2025splittingregions}.

\item \textbf{Special role of equal-mass exchange:}
For equal-mass exchanges ($s_{12}=s_{34}$) the kinematics degenerates and the partial-wave expansion collapses to the diagonal $j=l$ sector, with the relative $\omega$-harmonic weights fixed by spinor-helicity gluing.
This corner provides a particularly tractable laboratory for testing additional constraints and may serve as a useful intermediate step toward more general higher-spin and higher-point analyses.

\item \textbf{Dispersion relations:}
Another natural direction to explore would be in the context of crossing-symmetric dispersion relations, which have now been established firmly for the 2-2 case \cite{SDR, zahed, EE, venSDR}. In any dispersion relation, one standard approach is to utilize the partial wave expansion of the absorptive part of the amplitude - which has been the main focus of our work. Our analysis here will be useful for future investigations for the 5-particle dispersion relation, which was briefly examined in \cite{SDR}.

\end{itemize}

\section*{Acknowledgments} We thank Debapriyo Chowdhury, Soumen Saren, Athira P V and especially Apratim Kaviraj for discussions. APS would like to thank the participants of “Aspects of CFT-2" workshop held in Indian Institute of Technology, Kanpur in October, 2025 and “Progress of Theoretical Bootstrap” workshop held in Yukawa Institute for Theoretical Physics at Kyoto University in November, 2025, where some preliminary results were presented. ChatGPT-5.2 has been used to review the literature, check grammar and to perform some of the algebra (we have cross-checked the results). AS acknowledges support from an ANRF-ARG grant ANRF/ARG/2025/001338/PS and from a QHA senior fellowship. APS is supported by the DST INSPIRE Faculty Fellowship (IFA22-PH 282).
	
	\appendix
	
	\section{Four-point kinematics}
    \label{App:4-pt}
	We review the four-point kinematics here. 
    
	\subsection{Parameterization of momenta}
	We take the example of $2 \rightarrow 2$ scattering involving identical scalars of mass $m$ in the external legs, where the incoming states are 1 and 2, and the outgoing states are 3 and 4. 
	\begin{center}
		\begin{tikzpicture}
			\draw [gray, thin] (-5,2) -- (-3,0);
			\draw [gray, thin] (-5,-2) -- (-3,0);
			\draw [gray, thick] (-3,0)--(-1,0);
			\draw [gray, thick] (1,0)--(3,0);
			\draw [gray, thin] (3,0) -- (5,2);
			\draw [gray, thin] (3,0) -- (5,-2);
			\node at (-5.1,2.1) {$1$};
			\node at (-5.1, -2.1) {$2$};
			\node at (5.1,-2.1) {$3$};
			\node at (5.1,2.1) {$4$};
			\node at (-2,-0.22) {$Q$};
			\node at (2,-0.22) {$Q$};
			\node at (-2,0.22) {$\rightarrow$};
			\node at (2,0.22) {$\rightarrow$};
		\end{tikzpicture}
	\end{center}
	Kinematics of the scattering process can be understood as a mapping between center of mass (COM) frames of the initial states and the final states. Momenta in the initial COM frame can be parameterized as 
	\begin{eqnarray}
		p_{1}^{(i)\mu} & = & \left(\frac{E}{2}, 0,0, \sqrt{\frac{E^{2}}{4}-m^{2}}\right)~, \nonumber\\
		p_{2}^{(i)\mu} & = &\left(\frac{E}{2}, 0,0, -\sqrt{\frac{E^{2}}{4}-m^{2}}\right)~, \nonumber\\
		Q^{(i)\mu} & = & \left(E, 0, 0, 0\right)~.
	\end{eqnarray}
	In the final COM frame the momenta are given by
	\begin{eqnarray}
		Q^{(f)\mu} & = & \left(E', 0, 0, 0\right)~,\nonumber\\
		p_{3}^{(f)\mu} & = & \left(\frac{E'}{2}, 0, 0, -\sqrt{\frac{E^{'2}}{4} - m^{2}}\right)~, \nonumber\\
		p_{4}^{(f)\mu} & = & \left(\frac{E'}{2}, 0, 0, \sqrt{\frac{E^{'2}}{4} - m^{2}}\right)~.
	\end{eqnarray}
	Conservation of momenta in each COM frame leads to $p_{1}+p_{2}=Q$.
	
	We can apply boosts and rotations to go from the initial to the final frames. Conservation of total momentum requires $E=E'$. This fixes the boost parameter. A general transformation involving rotations can be given as functions of Euler angles, 
	\begin{equation}
		R_{i\rightarrow f} = U_{z}\left(\phi\right)U_{x}\left(\theta\right)U_{z}\left(\psi\right).
	\end{equation}
	Momentum transfer between a final and an initial state can be calculated as follows, 
	\begin{eqnarray}
		t & = &  -\left(p_{1}-p_{4}\right)^{2}\nonumber\\
		& = & 2m^{2} + 2p_{1}^{(f)}\cdot p_{4}^{(f)}\nonumber\\
		& = & 2m^{2} + 2 \left[R_{i\rightarrow f}\;p_{1}^{(i)}\right]\cdot p_{4}^{(f)}~.
	\end{eqnarray}
	The last term in the above equation reads as 
	\begin{equation}
		\eta_{\mu\nu}\left(R_{i\rightarrow f}\right)^{\mu}_{\phantom{\mu}\sigma}p_{1}^{(i)\sigma}p_{4}^{(f)\mu}~.
	\end{equation}
	Rotation matrix along $x$-axis is 
	\begin{equation}
		U_{x}\left(\theta\right) = \begin{pmatrix}
			1 & 0 & 0 & 0 \\
			0 & 1 & 0 & 0 \\
			0 & 0 & \cos\theta & \sin\theta \\
			0 & 0 & -\sin\theta & \cos\theta  
		\end{pmatrix}~.
	\end{equation}
	Since $p_{1}$ and $p_{4}$ are oriented towards the $z$-axis in the respective frames, action of the rotation along $z$-axis is trivial. Putting the relevant expressions, finally we obtain
	\begin{equation}
		t = -\frac{s-4m^{2}}{2}\left(1-\cos\theta\right), \qquad \text{with}\: s=-\left(p_{1}+p_{2}\right)^{2}=E^{2}~.
	\end{equation}
	
	\subsection{Partial-waves for 4-point }
	Residue of the four-point amplitude over a factorization channel can be obtained by gluing two three-point amplitudes. For simplicity, let us assume the external states are massless scalars and a massive  particle with spin $j$ is being exchanged. Then the residue is given by \cite{Arkani-Hamed:2017jhn}
	\begin{eqnarray}
		\text{Res}_{s=m^{2}}\mathcal{M}\left(s,t\right) & = & \mathcal{A}_{L}\left(1,2,P^{(I_{1}\ldots I_{2j})}\right)\mathcal{A}_{R}\left(P_{(I_{1}\ldots I_{2j})},3,4\right)\nonumber\\
		& = & \frac{g^{2}}{m^{4j-2}}[12]^{j}[34]^{j}\langle1P^{(I_{1}}\rangle\ldots\langle1P^{I_{j}}\rangle\langle2P^{I_{j+1}}\rangle\ldots\langle2P^{I_{2j})}\rangle\nonumber\\
		&& \hspace{1cm}\times \langle P_{(I_{1}}3\rangle\ldots\langle P_{I_{j}}3\rangle\langle P_{I_{j+1}}4\rangle\ldots\langle P_{I_{2j})}4\rangle~.
	\end{eqnarray}
	Conventions for the spinor-helicity variables are noted in \ref{App:SH-notation}. Using the fact,
	\begin{eqnarray}
		\langle1P^{I}\rangle\langle P_{I}3\rangle = m\langle 13\rangle~, \quad 	\langle1P^{I}\rangle\langle P_{I}4\rangle = m\langle 14\rangle~,\nonumber\\
		\langle2P^{I}\rangle\langle P_{I}3\rangle = m\langle 23\rangle~, \quad 	\langle2P^{I}\rangle\langle P_{I}4\rangle = m\langle 24\rangle~,
	\end{eqnarray}
	and accounting for all possible contractions of $SU(2)$ indices, we get
	\begin{equation}
		\text{Res}_{s=m^{2}}\mathcal{M}\left(s,t\right)  =  \frac{g^{2}}{m^{2j-2}}[12]^{j}[34]^{j}\sum_{a=0}^{j}\left(\frac{j!}{a!\left(j-a\right)!}\right)^{2}\langle13\rangle^{a}\langle14\rangle^{j-a}\langle23\rangle^{j-a}\langle24\rangle^{a}
	\end{equation}
	Applying Lorentz transformations to spinor-heilicity variables, we obtain
	\begin{eqnarray}
		\langle13\rangle=-m\left(\frac{\cos\theta+1}{2}\right)^{\frac{1}{2}}~, \qquad \langle14\rangle = m\left(\frac{\cos\theta-1}{2}\right)^{\frac{1}{2}}~,\nonumber\\
		\langle23\rangle=-m\left(\frac{\cos\theta-1}{2}\right)^{\frac{1}{2}}~, \qquad \langle24\rangle=m\left(\frac{\cos\theta+1}{2}\right)^{\frac{1}{2}}~.
	\end{eqnarray}
	We also have  $[12]=-m$ and $[34]=m$. Finally, putting everything together, we get
	\begin{eqnarray}
		\text{Res}_{s=m^{2}}\mathcal{M}\left(s,t\right) & = & g^{2}m^{2j+2}\sum_{a=0}^{j}\left(\frac{j!}{a!\left(j-a\right)!}\right)^{2}\left(\frac{\cos\theta-1}{2}\right)^{j-a}\left(\frac{\cos\theta+1}{2}\right)^{a}\nonumber\\
		& = & g^{2}m^{2j+2} P_{j}\left(\cos\theta\right)~.
	\end{eqnarray}
	The last equality follows from the series expansion of Jacobi polynomials, $P_{n}^{(\alpha,\beta)}$ given by
	\begin{equation}
		P_{n}^{(\alpha,\beta)} \left(x\right)= \sum_{s=0}^{n}\begin{pmatrix}
			n+\alpha\\ n-s
		\end{pmatrix}\begin{pmatrix}
			n+\beta\\s
		\end{pmatrix}\left(\frac{x-1}{2}\right)^{s}\left(\frac{x+1}{2}\right)^{n-s}~.
	\end{equation} 
	Jacobi polynomials reduce to Legendre polynomials for $\alpha=0=\beta$.

	\section{Five-point kinematics}
	\label{app:5ptparameter}

    Here we consider the compatible set of factorization channels $\left(s_{12}, s_{34}\right)$ and work out the kinematics for the five-point tree-amplitude in this configuration.
	\begin{center}
		\begin{tikzpicture}[scale=0.7]
			\draw [gray, thick] (-3,2) -- (-1,0);
			\draw [gray, thick] (-3,-2) -- (-1,0);
			\draw [gray, thick] (-1,0) -- (4,0);
			\draw [gray, thick] (4,0) -- (6,2);
			\draw[gray, thick] (4,0) -- (6,-2);
			\draw [gray, thick] (1.5,0) -- (1.5,2);
			\node at (-3.5,2.5) {$1$};
			\node at (-3.5,-2.5) {$2$};
			\node at (6.5,-2.5) {$3$};
			\node at (6.5, 2.5) {$4$};
			\node at (1.5, 2.5) {$5$};
			\node at (0.5,-0.2) {$s_{12}$};
			\node at (2.5,-0.2) {$s_{34}$};
		\end{tikzpicture}
	\end{center}
    There are three three-point vertices in the above graph. Naively we may expect there are three center of mass frames associated to this configuration, however there are two frames corresponding to the middle three-point kinematic which are related to each other by a boost transformation.
	\begin{center}
		\begin{tikzpicture}[scale=0.7]
			\draw [gray, thin] (-10,2) -- (-8,0);
			\draw [gray, thin] (-10,-2) -- (-8,0);
			\draw [gray, thick] (-8,0)--(-6,0);
			\draw [gray, thick] (4,0)--(6,0);
			\draw [gray, thin] (6,0) -- (8,2);
			\draw [gray, thin] (6,0) -- (8,-2);
			\draw [gray, thin] (-1,2)--(-1,0);
			\draw [gray, thick](-3,0)--(-1,0);
			\draw [gray, thick] (-1,0)--(1,0);
			\node at (-10.1,2.1) {$1$};
			\node at (-10.1, -2.1) {$2$};
			\node at (8.1,-2.1) {$3$};
			\node at (8.1,2.1) {$4$};
			\node at (-1,2.2) {$5$};
			\node at (-7,-0.22) {$Q_{12}$};
			\node at (5,-0.22) {$Q_{34}$};
			\node at (-7,0.22) {$\rightarrow$};
			\node at (5,0.22) {$\rightarrow$};
			\node at (-2,-0.22) {$Q_{12}$};
			\node at (0,-0.22) {$Q_{34}$};
			\node at (-2,0.22) {$\rightarrow$};
			\node at (0,0.22) {$\rightarrow$};
		\end{tikzpicture}
	\end{center}
	The middle three-point kinematics can be analyzed from the COM frame of either $\left(p_{5}, Q_{12}\right)$ or $\left(p_{5}, Q_{34}\right)$. So we can assign four frames and the momenta are given by the following,
	\begin{itemize}
		\item[$\bullet$] Frame 1
		\begin{eqnarray}
			p_{1}^{(1)} & = & \left(\frac{\sqrt{s_{12}}}{2}, 0, 0, \sqrt{\frac{s_{12}}{4}-m^{2}}\right)~,\nonumber\\
			p_{2}^{(1)} & = & \left(\frac{\sqrt{s_{12}}}{2}, 0, 0, -\sqrt{\frac{s_{12}}{4}-m^{2}}\right)~,\nonumber\\
			Q_{12}^{(1)} & = & \left(\sqrt{s_{12}}, 0, 0, 0\right)~.
		\end{eqnarray} 
		Conservation of momenta implies $p_{1}^{(1)}+p_{2}^{(2)}=Q_{12}^{(1)}$.
		
		\item[$\bullet$] Frame 2
		\begin{eqnarray}
			Q_{12}^{(2)} & = & \left(\sqrt{s_{12}}, 0, 0, 0\right)~,\nonumber\\
			p_{5}^{(2)} & = & \left(E,0,0,\sqrt{E^{2}-m^{2}}\right)~,\nonumber\\
			Q_{34}^{(2)} & =  & \left(\sqrt{s_{12}}-E,0,0,-\sqrt{E^{2}-m^{2}}\right)~.
		\end{eqnarray}
		Conservation of momenta implies $Q_{12}^{(2)} = p_{5}^{(2)}+Q_{34}^{(2)}$.
		
		\item [$\bullet$] Frame 3
		\begin{eqnarray}
			Q_{12}^{(3)} & = & \left(\sqrt{s_{34}}+E',0,0,\sqrt{E^{'2}-m^{2}}\right)~,\nonumber\\
			p_{5}^{(3)} & = & \left(E',0,0,\sqrt{E^{'2}-m^{2}}\right)~,\nonumber\\
			Q_{34}^{(3)} & = & \left(\sqrt{s_{34}},0,0,0\right)~.
		\end{eqnarray}
		Conservation of momenta implies $Q_{12}^{(3)} = p_{5}^{(3)}+Q_{34}^{(3)}$. This is the COM frame for $Q_{12}^{(3)}$ and $-p_{5}^{(3)}$.
		
		\item [$\bullet$] Frame 4
		\begin{eqnarray}
			Q_{34}^{(4)} & = & \left(\sqrt{s_{34}},0,0,0\right)~, \nonumber\\
			p_{3}^{(4)} & = & \left(\frac{\sqrt{s_{34}}}{2}, 0, 0, -\sqrt{\frac{s_{34}}{4}-m^{2}}\right)~,\nonumber\\
			p_{4}^{(4)} & = & \left(\frac{\sqrt{s_{34}}}{2}, 0, 0, \sqrt{\frac{s_{34}}{4}-m^{2}}\right)
		\end{eqnarray}
		Conservation of momenta implies $Q_{34}^{(4)} = p_{3}^{(4)}+p_{4}^{(4)}$.
	\end{itemize}
	For the above parameterizations we used the conventions: $s_{ij}=-\left(p_{i}+p_{j}\right)^{2}$ and $p_{i}^{2}=-m^{2}$.
	
    The four center of mass frames are related to each other by the following transformations, 
	\begin{eqnarray}\label{Lor-transf}
		\mathcal{T}_{1\rightarrow 2} & = & U_{z}\left(\phi\right)U_{x}\left(\theta_{12}\right)U_{z}\left(\psi\right)~,\nonumber\\
		\mathcal{T}_{2\rightarrow 3} & = & \mathcal{B}_{z}\left(\zeta\right)~,\nonumber\\
		\mathcal{T}_{3\rightarrow 4} & = & U_{z}\left(\psi'\right)U_{x}\left(\theta_{34}\right)U_{z}\left(\phi'\right)~.
	\end{eqnarray}
	$U$ is the rotation along the corresponding axis. $\mathcal{B}_{z}\left(\zeta\right)$ is boost along $z$-axis and is given by
	\begin{equation}
		\mathcal{B}_{z}\left(\zeta\right) = \begin{pmatrix}
			\cosh\zeta & 0 & 0 & \sinh\zeta\\
			0 & 1 & 0 & 0 \\
			0 & 0 & 1 & 0 \\
			\sinh\zeta & 0 & 0 & \cosh\zeta
		\end{pmatrix}~.
	\end{equation}
	We get constraints from the transformations,
	\begin{equation}
		\left[\mathcal{B}_{z}\left(\zeta\right)\right]^{\mu}_{\phantom{\mu}\nu}\biggl\{p_{5}^{(2)\nu}, Q_{12}^{(2)\nu}, Q_{34}^{(2)\nu}\biggr\} = \biggl\{p_{5}^{(3)\mu}, Q_{12}^{(3)\mu}, Q_{34}^{(3)\mu}\biggr\}~.
	\end{equation}
	These constraints give the following solutions
	\begin{eqnarray}\label{transf-sol}
		E & = & \frac{s_{12}-s_{34}+m^{2}}{2\sqrt{s_{12}}}~,\nonumber\\
		E' & = & \frac{s_{12}-s_{34}-m^{2}}{2\sqrt{s_{34}}}~,\nonumber\\
		\zeta & = & \cosh^{-1}\left(\frac{s_{12}+s_{34}-m^{2}}{2\sqrt{s_{12}s_{34}}}\right)~.
	\end{eqnarray}
    Using \eqref{Lor-transf} we can then calculate $s_{23}$, $s_{45}$ and $s_{51}$. Expressions for these variables are given in \eqref{kinem-param}.
	Since momenta of the particles are aligned along the $z$-axis in every frame, actions of $U_{z}\left(\psi\right)$ and $U_{z}\left(\psi'\right)$ are trivial. Without loss of generality, we can set $\psi=0=\psi'$. Frame 1 is mapped to frame 4 by the convolution of three transformations,
	\begin{eqnarray}
		\mathcal{T}_{3\rightarrow 4}\mathcal{T}_{2\rightarrow 3}\mathcal{T}_{1\rightarrow 2} & = & U_{x}\left(\theta_{34}\right)U_{z}\left(\phi'\right)\mathcal{B}_{z}\left(\zeta\right)U_{z}\left(\phi\right)U_{x}\left(\theta_{12}\right)\nonumber\\
		& = & U_{x}\left(\theta_{34}\right)U_{z}\left(\phi+\phi'\right)\mathcal{B}_{z}\left(\zeta\right)U_{x}\left(\theta_{12}\right)~,
	\end{eqnarray}
	where the last line follows because rotation and boost along any axis commute. Denoting $\phi+\phi' = \omega$, we see that resultant transformation depends on the three angles $\left(\theta_{12}, \theta_{34},\omega\right)$. We will use $s_{12}$, $s_{34}$ and the three angles as a set of five independent variables to describe a five-point amplitude. 

    \subsection{Geometric meaning of the Toller angle at five points}
\label{sec:toller-angle}
In multiparticle scattering, $\omega$ is sometimes referred to as the Toller angle \cite{Brower:1974}. We will now give a geometric picture of the angles in the 5 particle case. 
For a double residue on $(s_{12},s_{34})$ it is natural to introduce the composite momenta
\begin{equation}
Q_{12}:=p_1+p_2,\qquad Q_{34}:=p_3+p_4,
\end{equation}
which meet particle $5$ at the central three-point vertex in the factorization channel.
A convenient geometric choice is to work in a frame where the direction of $p_5$ defines the
$\hat z$ axis.  Then $Q_{12}$ and $Q_{34}$ can be described by polar angles with respect to $p_5$,
\begin{equation}
\theta_{12} := \angle(Q_{12},p_5),\qquad \theta_{34}:=\angle(Q_{34},p_5),
\end{equation}
together with an azimuthal ``twist'' angle $\omega$ measuring the relative orientation of the two
planes
\begin{equation}
\Pi_{12}:=\mathrm{span}(p_5,Q_{12}),\qquad \Pi_{34}:=\mathrm{span}(p_5,Q_{34}).
\end{equation}
Equivalently, $\omega$ is the angle between the projections of $Q_{12}$ and $Q_{34}$ onto the plane
orthogonal to $p_5$.  With this definition one has the standard spherical-geometry identity
\begin{equation}
\widehat Q_{12}\!\cdot\!\widehat Q_{34}
=\cos\theta_{12}\cos\theta_{34}+\sin\theta_{12}\sin\theta_{34}\cos\omega,
\label{eq:toller-dot}
\end{equation}
which is precisely the origin of the characteristic $\sin\theta_{12}\sin\theta_{34}\cos\omega$ term
in the invariant relations (and hence in the residue), and explains why a Fourier expansion in
$e^{in\omega}$ is natural.

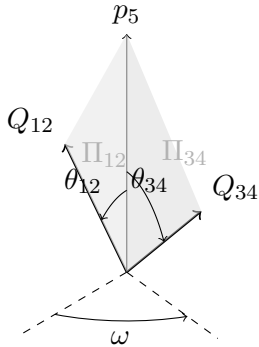
\begin{figure}[t]
\centering
\tdplotsetmaincoords{70}{120}
\begin{tikzpicture}[tdplot_main_coords, scale=1.05]
  \def\Lz{3.2}   
  \def\Lq{2.7}   
  \def\thA{35}   
  \def\thB{55}   
  \def\om{55}    
  \def\rA{1.10}  
  \def\rB{1.35}  

  \draw[->] (0,0,0) -- (0,0,\Lz) node[above] {$p_5$};

  \tdplotsetcoord{QA}{\Lq}{\thA}{0}
  \tdplotsetcoord{QB}{\Lq}{\thB}{\om}

  \draw[->, thick] (0,0,0) -- (QA) node[above left] {$Q_{12}$};
  \draw[->, thick] (0,0,0) -- (QB) node[above right] {$Q_{34}$};

  \fill[gray!15, opacity=0.70] (0,0,0) -- (0,0,\Lz) -- (QA) -- cycle;
  \fill[gray!30, opacity=0.50] (0,0,0) -- (0,0,\Lz) -- (QB) -- cycle;

  \draw[->]
    plot[domain=0:\thA, samples=40, variable=\a]
    ({\rA*sin(\a)*cos(0)}, {\rA*sin(\a)*sin(0)}, {\rA*cos(\a)});
  \node[anchor=south east] at ({\rA*sin(0.55*\thA)*cos(0)},
                               {\rA*sin(0.55*\thA)*sin(0)},
                               {\rA*cos(0.55*\thA)}) {$\theta_{12}$};

  \draw[->]
    plot[domain=0:\thB, samples=40, variable=\a]
    ({\rB*sin(\a)*cos(\om)}, {\rB*sin(\a)*sin(\om)}, {\rB*cos(\a)});
  \node[anchor=south] at ({\rB*sin(0.55*\thB)*cos(\om)},
                          {\rB*sin(0.55*\thB)*sin(\om)},
                          {\rB*cos(0.55*\thB)}) {$\theta_{34}$};

  \tdplotsetcoord{QAp}{\Lq}{90}{0}
  \tdplotsetcoord{QBp}{\Lq}{90}{\om}
  \draw[dashed] (0,0,0) -- (QAp);
  \draw[dashed] (0,0,0) -- (QBp);

  \tdplotdrawarc[->]{(0,0,0)}{1.8}{0}{\om}{anchor=north}{$\omega$}

  \node[gray!60] at ($(0,0,1.9)+(0.9,0.2,0)$) {$\Pi_{12}$};
  \node[gray!60] at ($(0,0,1.8)+(0.1,0.9,0)$) {$\Pi_{34}$};
\end{tikzpicture}
\caption{Toller (dihedral) angle for the five-point double-residue kinematics.  We define
$Q_{12}=p_1+p_2$ and $Q_{34}=p_3+p_4$ and choose a frame where $p_5$ sets the $\hat z$ axis.
The polar angles $\theta_{12}$ and $\theta_{34}$ are the angles between $Q_{12}$ and $p_5$, and
between $Q_{34}$ and $p_5$, respectively.  The Toller angle $\omega$ is the dihedral angle between
the planes $\Pi_{12}=\mathrm{span}(p_5,Q_{12})$ and $\Pi_{34}=\mathrm{span}(p_5,Q_{34})$, equivalently
the angle between the transverse projections of $Q_{12}$ and $Q_{34}$ onto the plane orthogonal to $p_5$.}
\label{fig:toller}
\end{figure}

    \section{Dictionary to classic multi-Regge kinematics}
    \label{App:MultiRegge}
A useful point of contact with the classic multi-Regge literature is the five-particle kinematic
parameterization of Brower--DeTar--Weis \cite{Brower:1974}, who introduce two polar angles and a
Toller (helicity) angle $\omega_{12}$ in a fixed coupling scheme.
In BDW notation one uses invariants
\[
t_1=Q_1^2=(p_a+p_{a'})^2,\qquad t_2=Q_2^2=(p_c+p_{c'})^2,\qquad
s_1=s_{a'b'}=(p_{a'}+p_{b'})^2,\qquad \] and \[
s_2=s_{b'c'}=(p_{b'}+p_{c'})^2,\qquad s_{12}=s_{ac}=(p_a+p_c)^2,
\]
together with angles $(\theta_1,\theta_2,\omega_{12})$, where $\omega_{12}$ is the Toller angle
whose Fourier modes carry Jacob--Wick helicity.
For our cyclic ordering $(1,2,3,4,5)$ and our choice of compatible channels $(s_{12},s_{34})$,
the identification
\[
(a',a,b',c,c')\equiv(1,2,5,3,4)
\]
implies the simple map
\[
t_1\leftrightarrow s_{12},\qquad t_2\leftrightarrow s_{34},\qquad
s_1\leftrightarrow s_{51},\qquad s_2\leftrightarrow s_{45},\qquad s_{12}\leftrightarrow s_{23},
\]
and $(\theta_1,\theta_2,\omega_{12})$ matches our three-angle variables
$(\theta_{12},\theta_{34},\omega)$ up to minor convention choices (e.g.\ $\theta\to\pi-\theta$ on one leg).
This makes it clear that our Fourier index $n$ in $e^{in\omega}$ is the finite-energy analogue of the
helicity label conjugate to the Toller angle in the classic partial-wave/multi-Regge analysis. 
While the kinematic backbone is the same, our use of partial waves differs from the classic multi-Regge treatment.  BDW formulate a $t$-channel partial-wave representation tailored to Regge and multi-Regge limits, with angular momentum in a given momentum-transfer channel and Fourier modes in the Toller angle encoding helicity \cite{Brower:1974}.  By contrast, we develop a finite-energy harmonic decomposition of \emph{double residues} on a compatible pair of poles (e.g.\ $s_{12}=m^2+k_1$ and $s_{34}=m^2+k_2$), expanding the residue as a function on the associated three-angle space in a basis of Gegenbauer/associated Legendre polynomials in $(x,y)$ together with Fourier modes $e^{in\omega}$.  This choice is designed to yield explicit inversion formulas for the coefficients $a^{(k_1,k_2)}_{jln}$ and to make splitting/hidden-zero constraints algebraic at the level of partial-wave data, rather than to analyze asymptotic Regge behavior.

	\section{Spinor helicity calculations}
    \label{App:SH-param}
	The validity of the basis for 5-point amplitude can be checked by gluing three-point amplitudes. In four dimension, spinor helicity variables provide a convenient method for constructing on-shell amplitudes from three-point functions. 
	
	\subsection{Notations}
    \label{App:SH-notation}
	Any four-momentum can be expressed in the following way,
	\begin{equation}
		p_{\alpha\dot{\alpha}} = \eta_{\mu\nu}p^{\mu}\sigma^{\nu}_{\alpha\dot{\alpha}} = \begin{pmatrix}
			-p^{0}+p^{3} & p^{1}-i p^{2}\\
			p^{1}+i p^{2} & -p^{0}-p^{3}
		\end{pmatrix}~.
	\end{equation}
	In case of null momentum determinant of the matrix vanishes and rank of the matrix reduces to one. Then we have
	\begin{equation}
		p_{\alpha\dot{\alpha}} = \lambda_{\alpha}\tilde{\lambda}_{\dot{\alpha}}~.
	\end{equation}
	For example, if $p^{\mu} = E\left(1,0,0,1\right)$ then 
	\begin{equation}
		\lambda_{\alpha} = \sqrt{2E} \begin{pmatrix}
			0 \\ 1
		\end{pmatrix}, \qquad 
		\tilde{\lambda}_{\dot{\alpha}} = -\sqrt{2E}\left(0,1\right)~.
	\end{equation}
	In our convention, if $\lambda_{\alpha} = \begin{pmatrix}
		a \\ b
	\end{pmatrix}$ then $\tilde{\lambda}_{\dot{\alpha}} = -\left(a^{\ast}, b^{\ast}\right)$.
	
	For massive momentum, $P^{2}=-m^{2}$, we will use
	\begin{equation}
		P_{\alpha\dot{\alpha}} = \lambda^{I}_{\alpha}\tilde{\lambda}_{I,\dot{\alpha}}. \qquad \text{$I$ represents $SU(2)$ index}.
	\end{equation}
    The massive spinor-helicity variables satisfy 
    \begin{equation}
        \epsilon_{IJ}\langle\lambda^{I}\lambda^{J}\rangle=m~, \quad \epsilon_{IJ}[\lambda^{I}\lambda^{J}]=-m~.
    \end{equation}
	
	\subsection{Frame transformations}
	\label{app:sphl-param}
	In \ref{app:5ptparameter} parameterization of momenta for five-point kinematic is given. Here we will find the analogous transformation rules for the spinor helicity variables. We set external momenta to be massless. 
	
	\paragraph{Frame 1}
	
	\begin{eqnarray}
		p^{(1)\mu}_{1} = \frac{\sqrt{s_{12}}}{2}\left(1,0,0,1\right) & \Rightarrow & \lambda^{(1)}_{1,\alpha} = \sqrt[4]{s_{12}}\begin{pmatrix}
			0 \\ 1
		\end{pmatrix},
		\quad \tilde{\lambda}^{(1)}_{1,\dot{\alpha}} = -\sqrt[4]{s_{12}}\left(0,1\right) \nonumber\\
		p^{(1)\mu}_{2} = \frac{\sqrt{s_{12}}}{2}\left(1,0,0,-1\right) & \Rightarrow & \lambda^{(1)}_{2,\alpha} = \sqrt[4]{s_{12}}\begin{pmatrix}
			1\\0
		\end{pmatrix},
		\quad \tilde{\lambda}^{(1)}_{2,\dot{\alpha}} = -\sqrt[4]{s_{12}}\left(1,0\right)
	\end{eqnarray}
	
	\paragraph{Frame 2}
	
	\begin{equation}
		p^{(2)\mu}_{5} = \frac{s_{12}-s_{34}}{2\sqrt{s_{12}}}\left(1,0,0,1\right) \quad \Rightarrow \quad \lambda^{(2)}_{5,\alpha} = \sqrt{\frac{s_{12}-s_{34}}{\sqrt{s_{12}}}}\begin{pmatrix}
			0\\1
		\end{pmatrix},
		\quad \tilde{\lambda}^{(2)}_{5,\dot{\alpha}} = -\sqrt{\frac{s_{12}-s_{34}}{\sqrt{s_{12}}}}\left(0,1\right)
	\end{equation}
	
	\paragraph{Frame 3}
	
	\begin{equation}
		p^{(3)\mu}_{5} = \frac{s_{12}-s_{34}}{2\sqrt{s_{34}}}\left(1,0,0,1\right) \quad\Rightarrow\quad \lambda^{(3)}_{5,\alpha} = \sqrt{\frac{s_{12}-s_{34}}{\sqrt{s_{34}}}}\begin{pmatrix}
			0\\1
		\end{pmatrix},
		\quad \tilde{\lambda}^{(3)}_{5,\dot{\alpha}} =  -\sqrt{\frac{s_{12}-s_{34}}{\sqrt{s_{34}}}}\left(0,1\right)
	\end{equation}
	
	\paragraph{Frame 4}
	
	\begin{eqnarray}
		p^{(4)\mu}_{3} = \frac{\sqrt{s_{34}}}{2}\left(1,0,0,-1\right) & \Rightarrow & \lambda^{(4)}_{3,\alpha} = \sqrt[4]{s_{34}}\begin{pmatrix}
			1\\0
		\end{pmatrix},
		\quad \tilde{\lambda}^{(4)}_{3,\dot{\alpha}}=-\sqrt[4]{s_{34}}\left(1,0\right) \nonumber\\
		p^{(4)\mu}_{4} = \frac{\sqrt{s_{34}}}{2}\left(1,0,0,1\right) &\Rightarrow & \lambda^{(4)}_{4,\alpha} = \sqrt[4]{s_{34}}\begin{pmatrix}
			0\\1
		\end{pmatrix},
		\quad \tilde{\lambda}^{(4)}_{4,\dot{\alpha}} = -\sqrt[4]{s_{34}}\left(0,1\right)
	\end{eqnarray}
	
	\paragraph{Lorentz transformations}
	Four-momenta transform under $SO(1,3)$ whereas the spinor-helicity variables will transform under $SL(2,C)$. Transformations are given by
	\begin{eqnarray}
		U\left(\mathbf{\theta}\right) \rightarrow e^{i\frac{\theta}{2}\hat{n}\cdot\sigma} &=& \cos\frac{\theta}{2}\; \mathbb{I} + i\sin\frac{\theta}{2}\; \hat{n}\cdot\sigma \qquad \text{(rotation)} \nonumber\\
		U\left(\mathbf{\zeta}\right) \rightarrow e^{\frac{\zeta}{2}\hat{n}\cdot\sigma} & = & \cosh\frac{\zeta}{2}\; \mathbb{I} + \sinh\frac{\zeta}{2}\; \hat{n}\cdot\sigma \qquad \text{(boost)}
	\end{eqnarray}
	Using the above equations, Lorentz transformations of \eqref{Lor-transf} acting on $\lambda$ can be found out to be 
	\begin{eqnarray}\label{spinor-transf}
		\mathcal{U}_{1\rightarrow 2} & = & \begin{pmatrix}
			\cos\frac{\theta_{12}}{2}e^{\frac{i}{2}\left(\phi+\psi\right)} & i \sin\frac{\theta_{12}}{2}e^{\frac{i}{2}\left(\phi-\psi\right)}\\
			i \sin\frac{\theta_{12}}{2}e^{-\frac{i}{2}\left(\phi-\psi\right)} & \cos\frac{\theta_{12}}{2}e^{-\frac{i}{2}\left(\phi+\psi\right)}
		\end{pmatrix}~,\nonumber\\
		\mathcal{U}_{2\rightarrow 3} & = & \begin{pmatrix}
			\sqrt[4]{\frac{s_{34}}{s_{12}}} & 0 \\
			0 & \sqrt[4]{\frac{s_{12}}{s_{34}}}
		\end{pmatrix}~, \nonumber\\
		\mathcal{U}_{3\rightarrow 4}& = & \begin{pmatrix}
			\sin \frac{\theta_{34}}{2}
			e^{\frac{i}{2}  \left(\phi'+\psi'\right)}
			& i \cos \frac{\theta_{34}}{2}
			e^{-\frac{i}{2}  \left(\phi'-\psi'\right)}
			\\
			i \cos \frac{\theta_{34}}{2}
			e^{\frac{i}{2}  \left(\phi'-\psi'\right)}
			& \sin \frac{\theta_{34}}{2}
			e^{-\frac{i}{2}  \left(\phi'+\psi'\right)}
		\end{pmatrix}~.
	\end{eqnarray}
	Note that in the last equality, we have changed $\theta_{34}\rightarrow \pi-\theta_{34}$.

\section{Physical region in terms of dimensionless invariants $\hat s_{ij}$}
\label{sec:physical-region}

We work with dimensionless Mandelstam invariants
\begin{equation}
\hat s_{ij}:=\frac{s_{ij}}{m^2},
\end{equation}
and treat $\{\hat s_{12},\hat s_{34},\hat s_{51},\hat s_{45},\hat s_{23}\}$ as independent variables, subject to the existence of real angles
$\theta_{12},\theta_{34},\omega$ with $\cos\theta_{12},\cos\theta_{34},\cos\omega\in[-1,1]$.
Introduce the standard velocity factors and K\"all\'en combination
\begin{equation}
\hat\beta_{12}:=\sqrt{1-\frac{4}{\hat s_{12}}},\qquad
\hat\beta_{34}:=\sqrt{1-\frac{4}{\hat s_{34}}},
\qquad
\hat\lambda:=(\hat s_{12}+\hat s_{34}-1)^2-4\hat s_{12}\hat s_{34},
\qquad
\hat\Delta:=\sqrt{\hat\lambda}.
\end{equation}
The physical region is characterized by the following necessary and sufficient inequalities.

\paragraph{(i) Reality of square-roots.}
Requiring $\hat\beta_{12},\hat\beta_{34},\hat\Delta\in\mathbb{R}$ gives
\begin{equation}
\boxed{\;
\hat s_{12}\ge 4,\qquad
\hat s_{34}\ge 4,\qquad
\hat\lambda\ge 0.
\;}
\label{eq:basic-real}
\end{equation}
On the usual $2\to 3$ (``physical'') branch where $\hat s_{12}$ is the total c.m.\ energy-squared of the incoming pair,
$\hat\lambda\ge 0$ is equivalent to
\begin{equation}
\boxed{\;
4\le \hat s_{34}\le\big(\sqrt{\hat s_{12}}-1\big)^2,
\;}
\label{eq:physical-branch}
\end{equation}
which is the standard threshold/phase-space bound.

\paragraph{(ii) Bounds from $|\cos\theta_{12}|\le 1$ and $|\cos\theta_{34}|\le 1$.}
From the compact relations
\begin{equation}
\hat s_{51}=\frac{3-\hat s_{12}+\hat s_{34}}{2}+\frac{\hat\Delta}{2}\,\hat\beta_{12}\cos\theta_{12},
\qquad
\hat s_{45}=\frac{3+\hat s_{12}-\hat s_{34}}{2}-\frac{\hat\Delta}{2}\,\hat\beta_{34}\cos\theta_{34},
\end{equation}
one finds
\begin{equation}
\cos\theta_{12}=\frac{2\hat s_{51}-(3-\hat s_{12}+\hat s_{34})}{\hat\Delta\,\hat\beta_{12}},
\qquad
\cos\theta_{34}=\frac{(3+\hat s_{12}-\hat s_{34})-2\hat s_{45}}{\hat\Delta\,\hat\beta_{34}}.
\end{equation}
Thus the constraints $|\cos\theta_{12}|\le 1$ and $|\cos\theta_{34}|\le 1$ are equivalent to
\begin{align}
\boxed{\;
\left|2\hat s_{51}-(3-\hat s_{12}+\hat s_{34})\right|
\le \hat\Delta\,\hat\beta_{12},
\;}
\label{eq:cos12-ineq}
\\
\boxed{\;
\left|(3+\hat s_{12}-\hat s_{34})-2\hat s_{45}\right|
\le \hat\Delta\,\hat\beta_{34}.
\;}
\label{eq:cos34-ineq}
\end{align}
Equivalently, for fixed $(\hat s_{12},\hat s_{34})$ the allowed ranges of $\hat s_{51}$ and $\hat s_{45}$ are the intervals
\begin{align}
\hat s_{51}
&\in
\Big[\tfrac{3-\hat s_{12}+\hat s_{34}}{2}-\tfrac{\hat\Delta}{2}\hat\beta_{12}\;,\;
     \tfrac{3-\hat s_{12}+\hat s_{34}}{2}+\tfrac{\hat\Delta}{2}\hat\beta_{12}\Big],
\label{eq:s51-interval}\\
\hat s_{45}
&\in
\Big[\tfrac{3+\hat s_{12}-\hat s_{34}}{2}-\tfrac{\hat\Delta}{2}\hat\beta_{34}\;,\;
     \tfrac{3+\hat s_{12}-\hat s_{34}}{2}+\tfrac{\hat\Delta}{2}\hat\beta_{34}\Big].
\label{eq:s45-interval}
\end{align}

\paragraph{(iii) The remaining bound $|\cos\omega|\le 1$ as an $\hat s_{23}$ band.}
Write the $\hat s_{23}$ relation in the form
\begin{equation}
\hat s_{23}=\hat s_{23}^{(0)}-\mathcal{B}\,\cos\omega,
\label{eq:s23-bandform}
\end{equation}
where
\begin{equation}
\mathcal{B}
=\frac{1}{2}\sqrt{\hat s_{12}\hat s_{34}}\;\hat\beta_{12}\hat\beta_{34}\;
\sin\theta_{12}\sin\theta_{34},
\qquad
\sin\theta_{12}=\sqrt{1-\cos^2\theta_{12}},\quad
\sin\theta_{34}=\sqrt{1-\cos^2\theta_{34}}.
\end{equation}
Using the identity
\begin{equation}
\hat\Delta\big(\hat\beta_{12}\cos\theta_{12}-\hat\beta_{34}\cos\theta_{34}\big)
=2(\hat s_{51}+\hat s_{45})-6,
\end{equation}
one convenient expression for the ``center'' $\hat s_{23}^{(0)}$ is
\begin{equation}
\hat s_{23}^{(0)}
=\frac{3-\hat s_{12}-\hat s_{34}}{4}+\frac{\hat s_{51}+\hat s_{45}}{2}
+\frac{\hat s_{12}+\hat s_{34}-1}{4}\,\hat\beta_{12}\hat\beta_{34}\,\cos\theta_{12}\cos\theta_{34}.
\end{equation}
Then $|\cos\omega|\le 1$ is equivalent to the band constraint
\begin{equation}
\boxed{\;
(\hat s_{23}-\hat s_{23}^{(0)})^2\le \mathcal{B}^2
\quad\Longleftrightarrow\quad
\hat s_{23}\in[\hat s_{23}^{(0)}-\mathcal{B},\;\hat s_{23}^{(0)}+\mathcal{B}] .
\;}
\label{eq:omega-band}
\end{equation}
Substituting \eqref{eq:cos12-ineq}--\eqref{eq:cos34-ineq} for $\cos\theta_{12},\cos\theta_{34}$ and
$\sin^2=1-\cos^2$ yields a purely algebraic (angle-free) description of the allowed region in terms of the five invariants.

\begin{figure}[htb]
  \centering
  \includegraphics[width=0.52\textwidth]{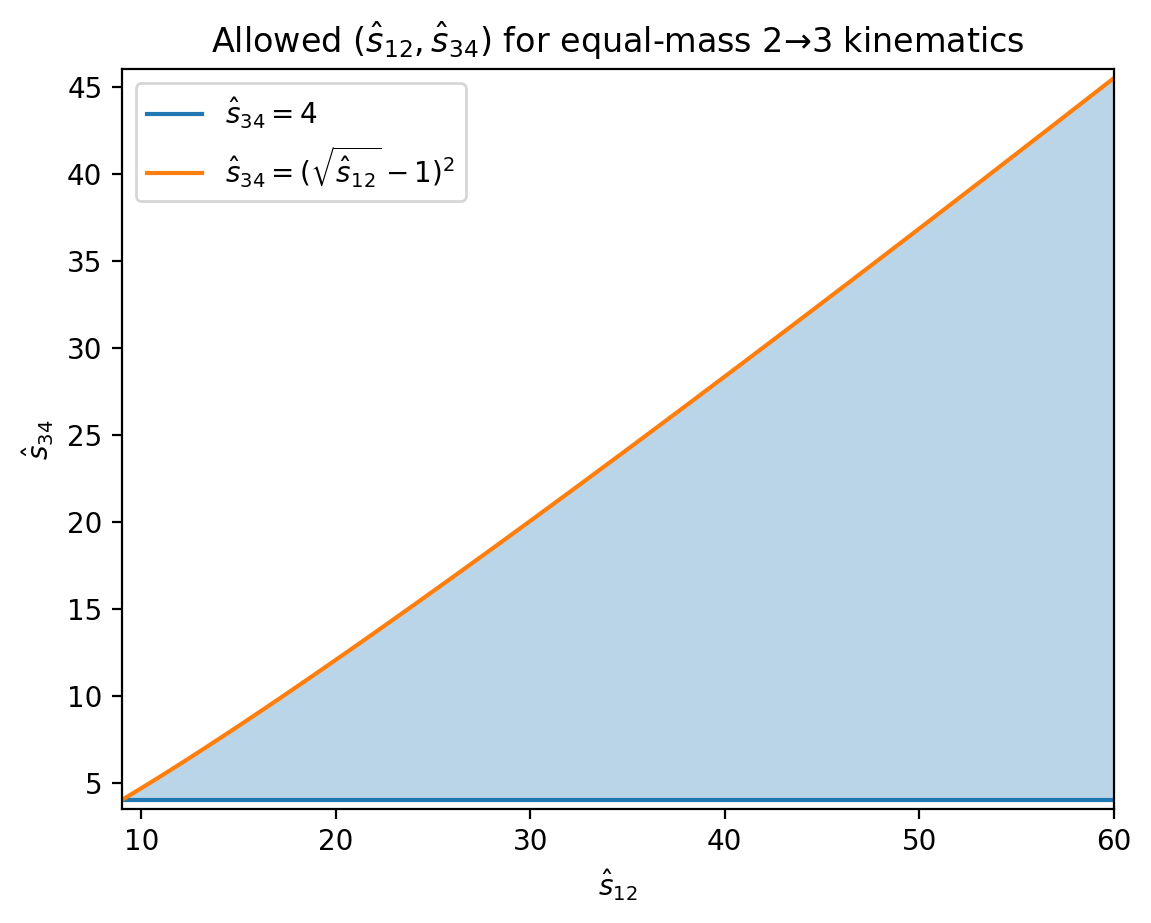}
  \caption{Allowed region in the $(\hat s_{12},\hat s_{34})$ plane for equal-mass $2\to3$ kinematics. The lower boundary is $\hat s_{34}=4$ and the upper boundary is $\hat s_{34}=(\sqrt{\hat s_{12}}-1)^2$, corresponding to $\hat\lambda\ge 0$ on the physical branch.}
  \label{fig:s12s34region}
\end{figure}

\begin{figure}[htb]
  \centering
  \includegraphics[width=0.62\textwidth]{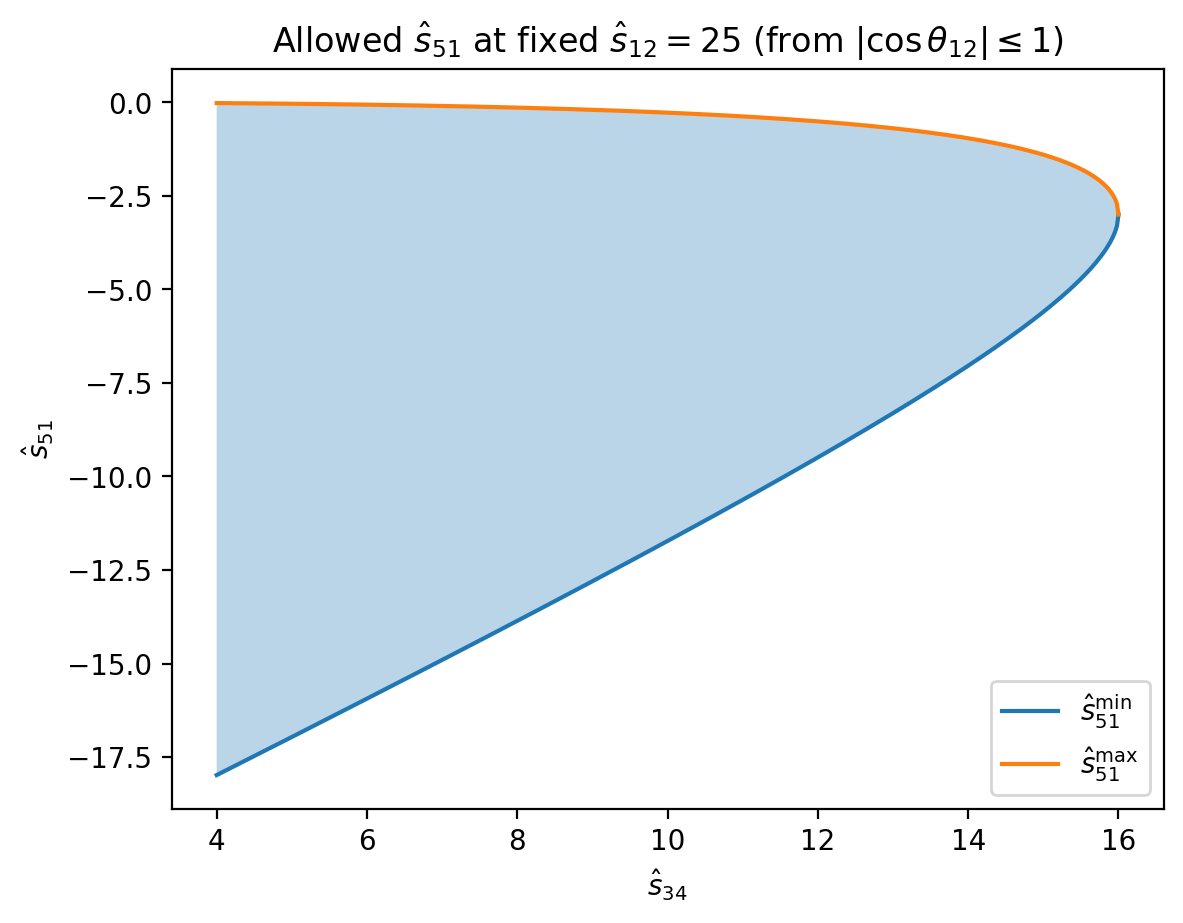}
  \caption{At fixed $\hat s_{12}=25$, the constraint $|\cos\theta_{12}|\le 1$ implies that $\hat s_{51}$ must lie in the interval \eqref{eq:s51-interval} as $\hat s_{34}$ varies over its allowed range.}
  \label{fig:s51band}
\end{figure}

\begin{figure}[htb]
  \centering
  \includegraphics[width=0.62\textwidth]{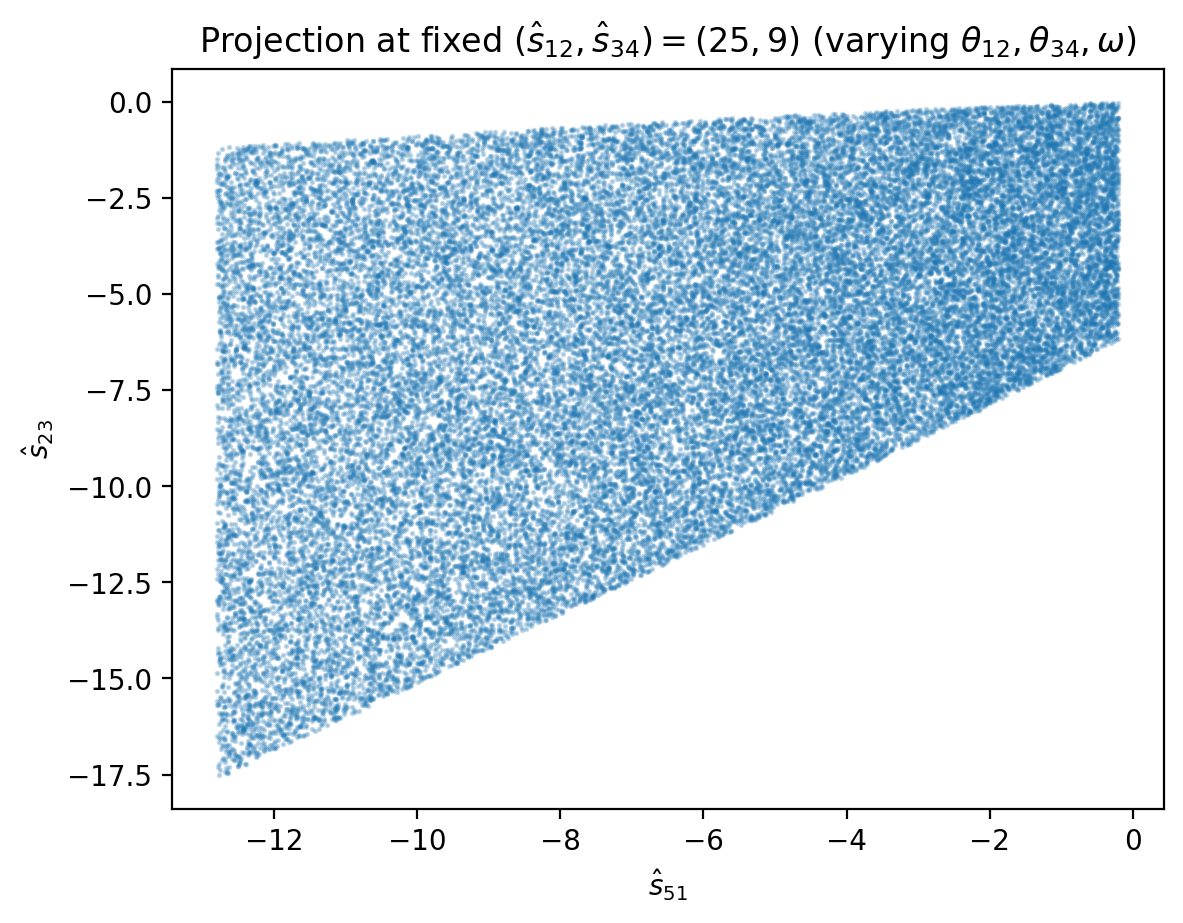}
  \caption{Projection of the allowed physical region at fixed $(\hat s_{12},\hat s_{34})=(25,9)$ into the $(\hat s_{51},\hat s_{23})$ plane, obtained by sampling $\theta_{12},\theta_{34},\omega$ over their full ranges. The thickness in $\hat s_{23}$ reflects the residual $\omega$ dependence encoded by the band \eqref{eq:omega-band}.}
  \label{fig:proj-s51-s23}
\end{figure}

\paragraph{Illustrations.}
Figure~\ref{fig:s12s34region} shows the allowed $(\hat s_{12},\hat s_{34})$ region for equal-mass $2\to 3$ kinematics.
Figure~\ref{fig:s51band} shows the interval \eqref{eq:s51-interval} for $\hat s_{51}$ as a function of $\hat s_{34}$ at fixed $\hat s_{12}=25$.
Finally, Figure~\ref{fig:proj-s51-s23} illustrates a projection of the full angular domain at fixed $(\hat s_{12},\hat s_{34})=(25,9)$ into the $(\hat s_{51},\hat s_{23})$ plane
(obtained by sampling $\theta_{12},\theta_{34},\omega$ uniformly).

\section{Lengthy splitting algebra}
\label{app:splitting-algebra}

This appendix collects some detailed intermediate steps of the splitting analyses presented in Sec.(\ref{Sec:splitting}). 

\subsection{Massive $(1,1)$ mass-level equations and intermediate relations}
\label{app:splitting-11}

On the locus $s_{13}=2m^{2}$ we have  $s_{45}=s_{12}+s_{23}-m^{2}$. On the pole $s_{12}=m^{2}+1$, splitting condition gives the following relation,
	 \begin{eqnarray}\label{constr1-11}
	 	&&\frac{\sqrt{m^2+1} a^{(1,1)}_{010} \left(-3 m^2+2
	 		s_{23}+2\right)}{\sqrt{-m^2} \sqrt{4-9
	 			\left(m^4+m^2\right)}}+\frac{\left(m^2+1\right)
	 		a^{(1,1)}_{110} \left(3 m^2-2
	 		\left(s_{23}+1\right)\right) \left(3 m^2-2
	 		s_{51}\right)}{m^2 \left(9
	 		\left(m^4+m^2\right)-4\right)} \nonumber\\
	 		&&+\frac{\sqrt{m^2+1}
	 		a^{(1,1)}_{100} \left(2 s_{51}-3
	 		m^2\right)}{\sqrt{-m^2} \sqrt{4-9
	 			\left(m^4+m^2\right)}} \nonumber\\
 			&&-\frac{2 a^{(1,1)}_{111}
	 		\left(3 m^6+m^4 \left(4-3 s_{23}\right)+m^2
	 		\left(s_{23} \left(s_{51}-5\right)-3\right)+2
	 		\left(s_{23}+1\right) s_{51}\right)}{m^2 \left(9
	 		\left(m^4+m^2\right)-4\right)}\nonumber\\
	 		&&+a^{(1,1)}_{000}-\left(\frac{2 c^{(1)}_{1} s_{23}}{1-3m^2}+c^{(1)}_{0}+c^{(1)}_{1}\right)
	 	\left(\frac{2 c^{(1)}_{1} s_{51}}{1-3
	 		m^2}+c^{(1)}_{0}+c^{(1)}_{1}\right) \nonumber\\
	 		& = & 0~.
	 \end{eqnarray}
	 Again, on the locus $s_{14}=2m^{2}$, we have the splitting condition after substituting $s_{23}=s_{45}+s_{51}-m^{2}$ on the five-point residue. Then we obatin the following condition,
	 \begin{eqnarray}\label{constr2-11}
	 	&& \frac{\sqrt{m^2+1} a^{(1,1)}_{010} \left(2 s_{45}-3 m^2\right)}{\sqrt{-m^2} \sqrt{4-9
	 			\left(m^4+m^2\right)}}+\frac{\left(m^2+1\right) a^{(1,1)}_{110} \left(3 m^2-2
	 		s_{45}\right) \left(3 m^2-2 s_{51}\right)}{m^2 \left(9
	 		\left(m^4+m^2\right)-4\right)}\nonumber\\
	 		&+& \frac{\sqrt{m^2+1} a^{(1,1)}_{100} \left(2 s_{51}-3
	 		m^2\right)}{\sqrt{-m^2} \sqrt{4-9 \left(m^4+m^2\right)}}\nonumber\\
	 		&-& \frac{2 a^{(1,1)}_{111}
	 		\left(6 m^6+m^4 \left(-3 s_{45}-3 s_{51}+8\right)+m^2 \left(s_{45}
	 		\left(s_{51}-5\right)-5 s_{51}-2\right)+2 s_{45} s_{51}\right)}{m^2 \left(9
	 		\left(m^4+m^2\right)-4\right)}+a^{(1,1)}_{000}\nonumber\\
	 		&-& \left(\frac{2 c^{(1)}_{1}
	 		s_{45}}{1-3 m^2}+c^{(1)}_{0}+c^{(1)}_{1}\right) \left(\frac{2 c^{(1)}_{1}
	 		s_{51}}{1-3 m^2}+c^{(1)}_{0}+c^{(1)}_{1}\right)\nonumber\\
	 		& = & 0~.
	 \end{eqnarray}
	 Using \eqref{constr1-11} we obtain the following set of relations,
	 \begin{eqnarray}
	 	a^{(1,1)}_{000}&=& \frac{\left(m^2+1\right) a^{(1,1)}_{110}}{3
	 		m^2-1}+\frac{1}{\left(3 m^2-1\right)^2} \biggl\{-9 \left(c^{(1)}_{0}\right)^2 m^4-3 \left(c^{(1)}_{1}\right)^2
	 		m^4+6 \left(c^{(1)}(0)\right)^2 m^2\nonumber\\
	 		&&-4 \left(c^{(1)}_{1}\right)^2 m^2+6
	 		c^{(1)}_{0} c^{(1)}_{1}
	 		m^2-\left(c^{(1)}_{0}\right)^2-c^{(1)}(1)^2-2 c^{(1)}(0)
	 		c^{(1)}_{1}\biggr\}~,\label{rel1-22}\\
	 		a^{(1,1)}_{010} & =& \frac{1}{\sqrt{m^2+1} \left(m^2+2\right)
	 			\left(3 m^2-1\right)^2}\biggl\{\sqrt{-m^2} \sqrt{-9 m^4-9 m^2+4} \nonumber\\
	 			&& \biggl\{3
	 			\left(c^{(1)}_{1}\right)^2 m^4-3 c^{(1)}_{0} c^{(1)}_{1} m^4+5
	 			\left(c^{(1)}_{1}\right)^2 m^2-5 c^{(1)}_{0} c^{(1)}_{1} m^2 +2
	 			\left(c^{(1)}_{1}\right)^2 +2 c^{(1)}_{0}
	 			c^{(1)}(1)\biggr\}\biggr\}\nonumber\\
	 			&& -\frac{\sqrt{-m^2}
	 			\sqrt{m^2+1} \sqrt{-9 m^4-9 m^2+4}
	 			a^{(1,1)}_{110}}{\left(m^2+2\right) \left(3
	 			m^2-1\right)}~,\label{rel2-22}\\
	 			a^{(1,1)}_{100} & = & \frac{\sqrt{-m^2} \sqrt{m^2+1} \sqrt{-9 m^4-9 m^2+4}
	 				a^{(1,1)}_{110}}{\left(m^2+2\right) \left(3
	 				m^2-1\right)}\nonumber\\
	 				&& -\frac{1}{\sqrt{m^2+1} \left(m^2+2\right)
	 				\left(3 m^2-1\right)^2}\biggl\{\sqrt{-m^2} \sqrt{-9 m^4-9
	 				m^2+4} \biggl\{3 \left(c^{(1)}_{1}\right)^2 m^4\nonumber\\
	 				&&+3 c^{(1)}_{0}
	 				c^{(1)}_{1} m^4+5 \left(c^{(1)}_{1}\right)^2 m^2+5 c^{(1)}_{0}
	 				c^{(1)}_{1} m^2+2 \left(c^{(1)}_{1}\right)^2-2 c^{(1)}_{0}
	 				c^{(1)}_{1}\biggr\}\biggr\}~,\label{rel3-22}\\
	 				a^{(1,1)}_{111} & = & \frac{2 \left(m^2+1\right)
	 					a^{(1,1)}_{110}}{m^2+2}-\frac{2 \left(c^{(1)}_{1}\right)^2 m^2
	 					\left(3 m^2+4\right)}{\left(m^2+2\right) \left(3
	 					m^2-1\right)}~.\label{rel4-22}
	 \end{eqnarray}
	 Again, from \eqref{constr2-11} we get two more relations,
	 \begin{eqnarray}
	 	a^{(1,1)}_{000} & = & -\frac{\left(m^2+1\right) a^{(1,1)}_{110}}{3
	 		m^2-1}-\frac{1}{\left(3 m^2-1\right)^2}\biggl\{-9 \left(c^{(1)}_{0}\right)^2 m^4-3 \left(c^{(1)}_{1}\right)^2
	 		m^4+6 \left(c^{(1)}_{0}\right)^2 m^2\nonumber\\
	 		&& -4 \left(c^{(1)}_{1}\right)^2 m^2+6
	 		c^{(1)}_{0} c^{(1)}_{1}
	 		m^2-\left(c^{(1)}_{0}\right)^2-\left(c^{(1)}_{1}\right)^2-2 c^{(1)}_{0}
	 		c^{(1)}_{1}\biggr\}~, \label{rel5-22}\\
	 		a^{(1,1)}_{100} & = & \frac{1}{\sqrt{m^2+1} \left(m^2+2\right)
	 			\left(3 m^2-1\right)^2} \biggl\{\sqrt{-m^2} \sqrt{-9 m^4-9 m^2+4} \nonumber\\
	 			&& \biggl\{3
	 			\left(c^{(1)}_{1}\right)^2 m^4-3 c^{(1)}_{0} c^{(1)}_{1} m^4+5
	 			\left(c^{(1)}_{1}\right)^2 m^2-5 c^{(1)}_{0} c^{(1)}_{1} m^2+2
	 			\left(c^{(1)}_{1}\right)^2+2 c^{(1)}_{0}
	 			c^{(1)}_{1}\biggr\}\biggr\}\nonumber\\
	 			&& -\frac{\sqrt{-m^2}
	 			\sqrt{m^2+1} \sqrt{-9 m^4-9 m^2+4}
	 			a^{(1,1)}_{110}}{\left(m^2+2\right) \left(3
	 			m^2-1\right)}~.\label{rel6-22}
	 \end{eqnarray}
	 Now equating the expressions for $a^{(1,1)}_{000}$ in \eqref{rel1-22} and \eqref{rel5-22} along with that of $a^{(1,1)}_{100}$ in \eqref{rel3-22} and \eqref{rel6-22} we get 

\subsection{Explicit $(1,2)$ solution from splitting}
\label{app:splitting-12}

\begin{eqnarray}\label{split-12}
		a^{(1,2)}_{000} & = & \frac{c^{(1)}_0}{\left(2-3
			m^2\right)^2 \left(m^4+3 m^2+2\right)}\biggl\{ c^{(2)}_0 \left(m^4+5
			m^2+6\right) \left(2-3 m^2\right)^2\nonumber\\
			&& \hspace{1cm} +c^{(2)}_2
			\left(-6 m^6+m^4+27 m^2-2\right)\biggr\}~, \nonumber\\
			a^{(1,2)}_{010} & = & -\frac{c^{(1)}_0 c^{(2)}_1}{\sqrt{2-3 m^2} \left(m^2+1\right)
				\sqrt{m^2+2} \sqrt{-3 m^4-6 m^2+1}}\left(3 m^6+12 m^4+11
				m^2-2\right)~,\nonumber\\
				a^{(1,2)}_{020} & = & \frac{c^{(1)}_0 c^{(2)}_2 }{3 m^4+4 m^2-4}\left(3 m^4+6
				m^2-1\right)~,\nonumber\\
				a^{(1,2)}_{100} & = & \frac{\sqrt{1-3 m^2} \sqrt{-3 m^4-6 m^2+1}
					 }{\left(2-3 m^2\right)^2
					\left(m^2+1\right)^{3/2} \left(m^2+2\right)}c^{(1)}_0\biggl\{c^{(2)}_0 \left(m^2+2\right) \left(2-3
					m^2\right)^2\nonumber\\
					&& \hspace{1cm}+c^{(2)}_2 \left(6 m^4+11
					m^2-6\right)\biggr\}~, \nonumber\\
					a^{(1,2)}_{110} & = & \frac{c^{(1)}_0 c^{(2)}_1 }{\sqrt{1-3 m^2} \sqrt{2-3 m^2}
						\left(m^2+1\right)^{3/2} \sqrt{m^2+2}}\left(9 m^6+18
						m^4-m^2-2\right)~,\nonumber\\
						a^{(1,2)}_{111} & = & c^{(1)}_0 \biggl\{\frac{3 c^{(2)}_2 \sqrt{1-3 m^2}
							\left(3 m^2+5\right) \left(3 m^4+6
							m^2-1\right)}{\left(2-3 m^2\right)^{3/2}
							\left(m^2+3\right)^2} \nonumber\\
							&& +\frac{\left(3 m^2-1\right)
							\left(3 m^4+6 m^2-1\right) \left(2 c^{(2)}_1
							\left(3 m^4+7 m^2-6\right)+3 c^{(2)}_2 \left(3
							m^4+8 m^2+5\right)\right)}{\sqrt{\frac{1-3 m^2}{2-3
									m^2}} \left(m^2+1\right) \left(3 m^4+7
							m^2-6\right)^2} \nonumber\\
							&& +\frac{2 c^{(2)}_1 \left(9 m^6+18
							m^4-m^2-2\right)}{\sqrt{1-3 m^2} \sqrt{2-3 m^2}
							\left(m^4+4 m^2+3\right)}\biggr\}~, \nonumber\\
							a^{(1,2)}_{120} & = & \frac{ \sqrt{1-3 m^2} \left(3
								m^2+5\right) \sqrt{-3 m^4-6 m^2+1}}{\sqrt{m^2+1}
								\left(3 m^4+4 m^2-4\right)}c^{(1)}_0 c^{(2)}_2~, \nonumber\\
								a^{(1,2)}_{121} & = & \frac{2  \sqrt{1-3 m^2}
									\sqrt{-3 m^4-6 m^2+1}}{\sqrt{m^2+2} \left(3
									m^4+m^2-2\right)}c^{(1)}_0 c^{(2)}_2~.
	\end{eqnarray}

\subsection{Splitting relations at higher mass-levels}
\label{App:split}

\paragraph{$s_{12}=1$, $s_{34}=3$ :} All the eleven partial-wave coefficients for the five-point residue can be determined from the four-point ones by the following relations, 
\begin{align}\label{split-13}
    a^{(1,3)}_{000} & =  \frac{4}{9} c^{(1)}_0 \left(9 c^{(3)}_0-2 c^{(3)}_2\right)~, & \qquad
		a^{(1,3)}_{010} & =  \frac{2}{27} c^{(1)}_0 \left(27 c^{(3)}_1-7 c^{(3)}_3\right)~, \nonumber\\
		a^{(1,3)}_{020} & =  \frac{8 c^{(1)}_0 c^{(3)}_2}{9}~, & \qquad 
		a^{(1,3)}_{030} & =  \frac{8 c^{(1)}_0 c^{(3)}_3}{27}~, \nonumber\\
		a^{(1,3)}_{100} & =  \frac{2}{9} c^{(1)}_0 \left(9 c^{(3)}_0-5 c^{(3)}_2\right)~, & \qquad 
		a^{(1,3)}_{110} & =  -\frac{28}{27} c^{(1)}_0 c^{(3)}_3~, \nonumber\\
		a^{(1,3)}_{111} & =  \frac{2 c^{(1)}_0 \left(5 c^{(3)}_3-9 c^{(3)}_1\right)}{9 \sqrt{3}}~, & \qquad
		a^{(1,3)}_{120} & =  -\frac{8}{9} c^{(1)}_0 c^{(3)}_2~,\nonumber\\
		a^{(1,3)}_{121} & =  -\frac{4 c^{(1)}_0 c^{(3)}_2}{3 \sqrt{3}}~,& \qquad
		a^{(1,3)}_{130} & =  -\frac{32}{27} c^{(1)}_0 c^{(3)}_3~,\nonumber\\
		a^{(1,3)}_{131} & =  -\frac{8 c^{(1)}_0 c^{(3)}_3}{9 \sqrt{3}}~. & 
\end{align}

\paragraph{$s_{12}=2$, $s_{34}=3$ :} There are twenty coefficients for the five-point residue. Eighteen of them can be determined in terms of the remaining two coefficients and the four-point ones. We choose the undetermined coefficients to be $a^{(2,3)}_{000}$ and $a^{(2,3)}_{110}$. Following are the constraining relations, 
\begin{eqnarray}\label{split-23}
		a^{(2,3)}_{010} & = & \frac{1}{810} \left(3348 a^{(2,3)}_{110}+81 \left(98c^{(2)}_1-31c^{(2)}_0\right)
		c^{(3)}_1+4 \left(861c^{(2)}_0-655c^{(2)}_1\right) c^{(3)}_3\right)~,\nonumber\\
		a^{(2,3)}_{020} & = & \frac{1}{38} \left(-2 a^{(2,3)}_{000}+c^{(2)}_1 \left(9 c^{(3)}_0-5 c^{(3)}_2\right)-4
		c^{(2)}_0 \left(c^{(3)}_0-2 c^{(3)}_2\right)\right)~,\nonumber\\
		a^{(2,3)}_{030} & = & \frac{1}{810} \left(-108 a^{(2,3)}_{110}+81 \left(c^{(2)}_0-3c^{(2)}_1\right)
		c^{(3)}_1+\left(25c^{(2)}_1-24c^{(2)}_0\right) c^{(3)}_3\right)~,\nonumber\\
		a^{(2,3)}_{100} & = & \frac{1}{684} \left(270 a^{(2,3)}_{000}-999c^{(2)}_0 c^{(3)}_0+666c^{(2)}_1
		c^{(3)}_0+60c^{(2)}_0 c^{(3)}_2-484c^{(2)}_1 c^{(3)}_2\right)~,\nonumber\\
		a^{(2,3)}_{111} & = & -\frac{36 a^{(2,3)}_{110}+c^{(2)}_1 \left(291 c^{(3)}_1-100
			c^{(3)}_3\right)+c^{(2)}_0 \left(68 c^{(3)}_3-117 c^{(3)}_1\right)}{30
			\sqrt{6}}~,\nonumber\\
			a^{(2,3)}_{120} & = & \frac{1}{684} \left(-270 a^{(2,3)}_{000}+135 \left(9c^{(2)}_1-4c^{(2)}_0\right)
			c^{(3)}_0+\left(453c^{(2)}_0-1055c^{(2)}_1\right) c^{(3)}_2\right)~,\nonumber\\
			a^{(2,3)}_{121} & = & \frac{-54 a^{(2,3)}_{000}+27 \left(9c^{(2)}_1-4c^{(2)}_0\right) c^{(3)}_0+2
				\left(51c^{(2)}_0-115c^{(2)}_1\right) c^{(3)}_2}{114 \sqrt{6}}~,\nonumber\\
				a^{(2,3)}_{130} & = & \frac{1}{36} \left(-36 a^{(2,3)}_{110}+3c^{(2)}_0 \left(9 c^{(3)}_1-11
				c^{(3)}_3\right)+c^{(2)}_1 \left(19 c^{(3)}_3-81 c^{(3)}_1\right)\right)~,\nonumber\\
				a^{(2,3)}_{131} & = & \frac{-108 a^{(2,3)}_{110}+81 \left(c^{(2)}_0-3c^{(2)}_1\right) c^{(3)}_1+11
					\left(5c^{(2)}_1-9c^{(2)}_0\right) c^{(3)}_3}{135 \sqrt{6}}~,\nonumber\\
					a^{(2,3)}_{200} & = & \frac{1}{684} \left(18 a^{(2,3)}_{000}+45 \left(2c^{(2)}_1-3c^{(2)}_0\right)
					c^{(3)}_0+4 \left(39c^{(2)}_0-41c^{(2)}_1\right) c^{(3)}_2\right)~,\nonumber\\
					a^{(2,3)}_{210} & = & \frac{1}{810} \left(-1242 a^{(2,3)}_{110}+81 \left(19c^{(2)}_0-42c^{(2)}_1\right)
					c^{(3)}_1+4 \left(475c^{(2)}_1-429c^{(2)}_0\right) c^{(3)}_3\right)~,\nonumber\\
					a^{(2,3)}_{211} & = & \frac{-108 a^{(2,3)}_{110}+27 \left(5c^{(2)}_0-11c^{(2)}_1\right) c^{(3)}_1+4
						\left(43c^{(2)}_1-39c^{(2)}_0\right) c^{(3)}_3}{54 \sqrt{6}}~,\nonumber\\
						a^{(2,3)}_{220} & = & \frac{1}{684} \left(\left(213c^{(2)}_0+83c^{(2)}_1\right) c^{(3)}_2-333 \left(2
						a^{(2,3)}_{000}+\left(4c^{(2)}_0-9c^{(2)}_1\right) c^{(3)}_0\right)\right)~,\nonumber\\
						a^{(2,3)}_{221} & = & \frac{-90 a^{(2,3)}_{000}+18c^{(2)}_0 \left(c^{(3)}_2-10
							c^{(3)}_0\right)+c^{(2)}_1 \left(405 c^{(3)}_0+22 c^{(3)}_2\right)}{114
							\sqrt{6}}~,\nonumber\\
							a^{(2,3)}_{222} & = & \frac{1}{228} \left(-18 a^{(2,3)}_{000}-4c^{(2)}_0 \left(9
							c^{(3)}_0+c^{(3)}_2\right)+3c^{(2)}_1 \left(27 c^{(3)}_0+4
							c^{(3)}_2\right)\right)~,\nonumber\\
							a^{(2,3)}_{230} & = & \frac{-3996 a^{(2,3)}_{110}+2997 \left(c^{(2)}_0-3c^{(2)}_1\right)
								c^{(3)}_1+\left(5485c^{(2)}_1-4893c^{(2)}_0\right) c^{(3)}_3}{1620}~,\nonumber\\
								a^{(2,3)}_{231} & = & \frac{-108 a^{(2,3)}_{110}+81 \left(c^{(2)}_0-3c^{(2)}_1\right)
									c^{(3)}_1+\left(151c^{(2)}_1-135c^{(2)}_0\right) c^{(3)}_3}{81 \sqrt{6}}~,\nonumber\\
									a^{(2,3)}_{232} & = & -\frac{a^{(2,3)}_{110}}{15}+\frac{1}{20} \left(c^{(2)}_0-3c^{(2)}_1\right)
									c^{(3)}_1+\frac{4}{405} \left(10c^{(2)}_1-9c^{(2)}_0\right) c^{(3)}_3~.
	\end{eqnarray}

\bibliographystyle{utphys}
	\bibliography{Dispersion}

@article{Sinha:2020win,
	author = "Sinha, Aninda and Zahed, Ahmadullah",
	title = "{Crossing Symmetric Dispersion Relations in Quantum Field Theories}",
	eprint = "2012.04877",
	archivePrefix = "arXiv",
	primaryClass = "hep-th",
	doi = "10.1103/PhysRevLett.126.181601",
	journal = "Phys. Rev. Lett.",
	volume = "126",
	number = "18",
	pages = "181601",
	year = "2021"
}

@article{deRham:2022gfe,
	author = "de Rham, Claudia and Jaitly, Sumer and Tolley, Andrew J.",
	title = "{Constraints on Regge behavior from IR physics}",
	eprint = "2212.04975",
	archivePrefix = "arXiv",
	primaryClass = "hep-th",
	reportNumber = "Imperial/TP/2022/CdR/05",
	doi = "10.1103/PhysRevD.108.046011",
	journal = "Phys. Rev. D",
	volume = "108",
	number = "4",
	pages = "046011",
	year = "2023"
}

@article{Arkani-Hamed:2017jhn,
	author = "Arkani-Hamed, Nima and Huang, Tzu-Chen and Huang, Yu-tin",
	title = "{Scattering amplitudes for all masses and spins}",
	eprint = "1709.04891",
	archivePrefix = "arXiv",
	primaryClass = "hep-th",
	reportNumber = "NCTS-TH/1714, NCTS-TH-1714",
	doi = "10.1007/JHEP11(2021)070",
	journal = "JHEP",
	volume = "11",
	pages = "070",
	year = "2021"
}

@article{White:1971fz,
	author = "White, A. R.",
	title = "{Fixed-angle limits of multiparticle amplitudes}",
	doi = "10.1007/BF02731528",
	journal = "Nuovo Cim. A",
	volume = "4",
	pages = "957--982",
	year = "1971"
}

@article{White:1972sc,
	author = "White, A. R.",
	title = "{The signatured froissart-gribov continuation of multiparticle amplitudes to complex helicity and angular momentum}",
	doi = "10.1016/0550-3213(72)90381-1",
	journal = "Nucl. Phys. B",
	volume = "39",
	pages = "432--460",
	year = "1972"
}

@article{Arkani-Hamed:2024nzc,
	author = "Arkani-Hamed, Nima and Figueiredo, Carolina and Remmen, Grant N.",
	title = "{Open string amplitudes: singularities, asymptotics and new representations}",
	eprint = "2412.20639",
	archivePrefix = "arXiv",
	primaryClass = "hep-th",
	doi = "10.1007/JHEP04(2025)039",
	journal = "JHEP",
	volume = "04",
	pages = "039",
	year = "2025"
}

@article{Arkani-Hamed:2023swr,
	author = "Arkani-Hamed, Nima and Cao, Qu and Dong, Jin and Figueiredo, Carolina and He, Song",
	title = "{Hidden zeros for particle/string amplitudes and the unity of colored scalars, pions and gluons}",
	eprint = "2312.16282",
	archivePrefix = "arXiv",
	primaryClass = "hep-th",
	doi = "10.1007/JHEP10(2024)231",
	journal = "JHEP",
	volume = "10",
	pages = "231",
	year = "2024"
}

@article{Berman:2025splittingregions,
  author        = {Berman, Justin and Elvang, Henriette and Figueiredo, Carolina},
  title         = {Splitting Regions and Shrinking Islands from Higher Point Constraints},
  eprint        = {2506.22538},
  archivePrefix = {arXiv},
  primaryClass  = {hep-th},
  journal       = {JHEP},
  volume        = {10},
  pages         = {226},
  year          = {2025},
  doi           = {10.1007/JHEP10(2025)226}
}

@article{Kruczenski:2022,
  title         = {Snowmass White Paper: S-matrix Bootstrap},
  author        = {Kruczenski, Martin and Penedones, Jo{\~a}o and van Rees, Balt C.},
  eprint        = {2203.02421},
  archivePrefix = {arXiv},
  primaryClass  = {hep-th},
  year          = {2022}
}

@article{Haring:2023,
  author = {H{\"a}ring, Kelian and Zhiboedov, Alexander},
    title = "{The stringy S-matrix bootstrap: maximal spin and superpolynomial softness}",
    eprint = "2311.13631",
    archivePrefix = "arXiv",
    primaryClass = "hep-th",
    reportNumber = "CERN-TH-2023-214",
    doi = "10.1007/JHEP10(2024)075",
    journal = "JHEP",
    volume = "10",
    pages = "075",
    year = "2024"
}

@article{Cheung:2024Strings,
  author = "Cheung, Clifford and Hillman, Aaron and Remmen, Grant N.",
    title = "{Bootstrap Principle for the Spectrum and Scattering of Strings}",
    eprint = "2406.02665",
    archivePrefix = "arXiv",
    primaryClass = "hep-th",
    reportNumber = "CALT-TH 2024-022",
    doi = "10.1103/PhysRevLett.133.251601",
    journal = "Phys. Rev. Lett.",
    volume = "133",
    number = "25",
    pages = "251601",
    year = "2024"
}

@article{Cheung:2025Multipositivity,
  author = "Cheung, Clifford and Remmen, Grant N.",
    title = "{Multipositivity bounds for scattering amplitudes}",
    eprint = "2505.05553",
    archivePrefix = "arXiv",
    primaryClass = "hep-th",
    reportNumber = "CALT-TH 2025-010",
    doi = "10.1103/wt4x-2149",
    journal = "Phys. Rev. D",
    volume = "112",
    number = "1",
    pages = "016017",
    year = "2025"
}

@article{Henn:2024SixPointIntegrals,
  author = "Henn, Johannes M. and Matija{\v{s}}i{\'c}, Antonela and Miczajka, Julian and Peraro, Tiziano and Xu, Yingxuan and Zhang, Yang",
    title = "{A computation of two-loop six-point Feynman integrals in dimensional regularization}",
    eprint = "2403.19742",
    archivePrefix = "arXiv",
    primaryClass = "hep-ph",
    reportNumber = "MPP-2024-53, USTC-ICTS/PCFT-24-11",
    doi = "10.1007/JHEP08(2024)027",
    journal = "JHEP",
    volume = "08",
    pages = "027",
    year = "2024"
}

@book{Eden:1966Book,
  author    = {Eden, R. J. and Landshoff, P. V. and Olive, D. I. and Polkinghorne, J. C.},
  title     = {The Analytic S-Matrix},
  publisher = {Cambridge University Press},
  year      = {1966}
}

@article{Virasoro:1969,
  author  = {Virasoro, M. A.},
  title   = {Generalization of Veneziano's Formula for the Five-Point Function},
  journal = {Phys. Rev. Lett.},
  volume  = {22},
  pages   = {37--39},
  year    = {1969},
  doi     = {10.1103/PhysRevLett.22.37}
}

@article{Mandelstam:1969,
  author  = {Mandelstam, S.},
  title   = {Generalizations of the Veneziano and Virasoro Models},
  journal = {Phys. Rev.},
  volume  = {183},
  pages   = {1374},
  year    = {1969},
  doi     = {10.1103/PhysRev.183.1374}
}

@article{White:1973,
  author  = {White, A. R.},
  title   = {Signature, factorization and unitarity in multi-Regge theory: The five-point function},
  journal = {Nucl. Phys. B},
  volume  = {67},
  pages   = {189--231},
  year    = {1973},
  doi     = {10.1016/0550-3213(73)90325-8}
}

@article{Brower:1974,
    author = "Brower, R. C. and DeTar, Carleton E. and Weis, J. H.",
    title = "{Regge Theory for Multiparticle Amplitudes}",
    reportNumber = "MIT-CTP-395, CERN-TH-1817",
    doi = "10.1016/0370-1573(74)90012-X",
    journal = "Phys. Rept.",
    volume = "14",
    pages = "257",
    year = "1974"
}

@phdthesis{Steinmann:1960,
  author = {Steinmann, O.},
  title  = {\"Uber den Zusammenhang zwischen den Wightmanfunktionen und den retardierten Kommutatoren},
  school = {ETH Z\"urich},
  year   = {1960}
}

@article{Cahill:1973,
  author  = {Cahill, K. E.},
  title   = {Generalized Optical Theorems and Generalized Steinmann Relations},
  journal = {Phys. Rev. D},
  volume  = {8},
  pages   = {2714},
  year    = {1973},
  doi     = {10.1103/PhysRevD.8.2714}
}

@article{Bartsch:2024HiddenZeros,
  author = "Bartsch, Christoph and Brown, Taro V. and Kampf, Karol and Oktem, Umut and Paranjape, Shruti and Trnka, Jaroslav",
    title = "{Hidden amplitude zeros from the double-copy map}",
    eprint = "2403.10594",
    archivePrefix = "arXiv",
    primaryClass = "hep-th",
    doi = "10.1103/PhysRevD.111.045019",
    journal = "Phys. Rev. D",
    volume = "111",
    number = "4",
    pages = "045019",
    year = "2025"
}

@article{Abarbanel:1972ayr,
    author = "Abarbanel, H. D. I. and Schwimmer, A.",
    title = "{Analytic structure of multiparticle amplitudes in complex helicity}",
    doi = "10.1103/PhysRevD.6.3018",
    journal = "Phys. Rev. D",
    volume = "6",
    pages = "3018--3031",
    year = "1972"
}

@ARTICLE{2019arXiv190106711S,
       author = {{Smirnov}, A.},
        title = "{View on N-dimensional spherical harmonics from the quantum mechanical P{\"o}schl-Teller potential well}",
      journal = {arXiv e-prints},
         year = 2019,
        month = jan,
          eid = {arXiv:1901.06711},
        pages = {arXiv:1901.06711},
          doi = {10.48550/arXiv.1901.06711},
archivePrefix = {arXiv},
       eprint = {1901.06711},
 primaryClass = {math-ph},
       adsurl = {https://ui.adsabs.harvard.edu/abs/2019arXiv190106711S}
}

@article{SDR,
    author = "Bhat, Faizan and Saha, Arnab Priya and Sinha, Aninda",
    title = "{A stringy dispersion relation for field theory}",
    eprint = "2506.03862",
    archivePrefix = "arXiv",
    primaryClass = "hep-th",
    month = "6",
    year = "2025"
}

@article{EE,
    author = "Bhat, Faizan and Chowdhury, Debapriyo and Saha, Arnab Priya and Sinha, Aninda",
    title = "{Bootstrapping string models with entanglement minimization and machine learning}",
    eprint = "2409.18259",
    archivePrefix = "arXiv",
    primaryClass = "hep-th",
    doi = "10.1103/PhysRevD.111.066013",
    journal = "Phys. Rev. D",
    volume = "111",
    number = "6",
    pages = "066013",
    year = "2025"
}

@article{venSDR,
    author = "Saha, Arnab Priya and Sinha, Aninda",
    title = "{Field Theory Expansions of String Theory Amplitudes}",
    eprint = "2401.05733",
    archivePrefix = "arXiv",
    primaryClass = "hep-th",
    doi = "10.1103/PhysRevLett.132.221601",
    journal = "Phys. Rev. Lett.",
    volume = "132",
    number = "22",
    pages = "221601",
    year = "2024"
}

@article{zahed,
    author = {Elias Mir{\'o}, Joan and Guerrieri, Andrea and G{\"u}m{\"u}s, Mehmet As{\i}m and Zahed, Ahmadullah},
    title = "{A Geometric View on Crossing-Symmetric Dispersion Relations}",
    eprint = "2509.14170",
    archivePrefix = "arXiv",
    primaryClass = "hep-th",
    month = "9",
    year = "2025"
}

@article{Elvang:2013cua,
    author = "Elvang, Henriette and Huang, Yu-tin",
    title = "{Scattering Amplitudes}",
    eprint = "1308.1697",
    archivePrefix = "arXiv",
    primaryClass = "hep-th",
    month = "8",
    year = "2013"
}

@article{Liu:2020fgu,
    author = "Liu, Jin-Yu and You, Zhe-Ming",
    title = "{The supersymmetric spinning polynomial}",
    eprint = "2011.11299",
    archivePrefix = "arXiv",
    primaryClass = "hep-th",
    month = "11",
    year = "2020"
}

@article{KNBalasubramanian:2022sae,
    author = "K. N. Balasubramanian, Mahesh and Chakraborty, Kushal and Rudra, Arnab and Saha, Arnab Priya",
    title = "{On-shell supersymmetry and higher-spin amplitudes}",
    eprint = "2209.06446",
    archivePrefix = "arXiv",
    primaryClass = "hep-th",
    doi = "10.1007/JHEP06(2023)037",
    journal = "JHEP",
    volume = "06",
    pages = "037",
    year = "2023"
}

@article{Guerrieri:2021ivu,
    author = "Guerrieri, Andrea and Penedones, Joao and Vieira, Pedro",
    title = "{Where Is String Theory in the Space of Scattering Amplitudes?}",
    eprint = "2102.02847",
    archivePrefix = "arXiv",
    primaryClass = "hep-th",
    doi = "10.1103/PhysRevLett.127.081601",
    journal = "Phys. Rev. Lett.",
    volume = "127",
    number = "8",
    pages = "081601",
    year = "2021"
}

@article{apratim5,
    author = "Antunes, Ant{\'o}nio and Harris, Sebastian and Kaviraj, Apratim",
    title = "{Five points for the Polyakov Bootstrap}",
    eprint = "2508.05623",
    archivePrefix = "arXiv",
    primaryClass = "hep-th",
    month = "8",
    year = "2025"
}

@article{poland,
    author = "Poland, David and Prilepina, Valentina and Tadi{\'c}, Petar",
    title = "{The five-point bootstrap}",
    eprint = "2305.08914",
    archivePrefix = "arXiv",
    primaryClass = "hep-th",
    doi = "10.1007/JHEP10(2023)153",
    journal = "JHEP",
    volume = "10",
    pages = "153",
    year = "2023"
}

@article{apratim6,
    author = "Antunes, Ant{\'o}nio and Harris, Sebastian and Kaviraj, Apratim and Schomerus, Volker",
    title = "{Lining up a positive semi-definite six-point bootstrap}",
    eprint = "2312.11660",
    archivePrefix = "arXiv",
    primaryClass = "hep-th",
    reportNumber = "DESY-23-220",
    doi = "10.1007/JHEP06(2024)058",
    journal = "JHEP",
    volume = "06",
    pages = "058",
    year = "2024"
}

@article{deRham:2017avq,
    author = "de Rham, Claudia and Melville, Scott and Tolley, Andrew J. and Zhou, Shuang-Yong",
    title = "{Positivity bounds for scalar field theories}",
    eprint = "1702.06134",
    archivePrefix = "arXiv",
    primaryClass = "hep-th",
    doi = "10.1103/PhysRevD.96.081702",
    journal = "Phys. Rev. D",
    volume = "96",
    number = "8",
    pages = "081702",
    year = "2017"
}

@article{deRham:2017zjm,
    author = "de Rham, Claudia and Melville, Scott and Tolley, Andrew J. and Zhou, Shuang-Yong",
    title = "{UV complete me: Positivity Bounds for Particles with Spin}",
    eprint = "1706.02712",
    archivePrefix = "arXiv",
    primaryClass = "hep-th",
    doi = "10.1007/JHEP03(2018)011",
    journal = "JHEP",
    volume = "03",
    pages = "011",
    year = "2018"
}

@article{Tolley:2020gtv,
    author = "Tolley, Andrew J. and Wang, Zi-Yue and Zhou, Shuang-Yong",
    title = "{New positivity bounds from full crossing symmetry}",
    eprint = "2011.02400",
    archivePrefix = "arXiv",
    primaryClass = "hep-th",
    doi = "10.1007/JHEP05(2021)255",
    journal = "JHEP",
    volume = "05",
    pages = "255",
    year = "2021"
}

@article{Caron-Huot:2020cmc,
    author = "Caron-Huot, Simon and Van Duong, Vincent",
    title = "{Extremal Effective Field Theories}",
    eprint = "2011.02957",
    archivePrefix = "arXiv",
    primaryClass = "hep-th",
    doi = "10.1007/JHEP05(2021)280",
    journal = "JHEP",
    volume = "05",
    pages = "280",
    year = "2021"
}

@article{Arkani-Hamed:2017mur,
    author = "Arkani-Hamed, Nima and Bai, Yuntao and He, Song and Yan, Gongwang",
    title = "{Scattering Forms and the Positive Geometry of Kinematics, Color and the Worldsheet}",
    eprint = "1711.09102",
    archivePrefix = "arXiv",
    primaryClass = "hep-th",
    doi = "10.1007/JHEP05(2018)096",
    journal = "JHEP",
    volume = "05",
    pages = "096",
    year = "2018"
}

@article{Bercini:2020msp,
    author = "Bercini, Carlos and Gon{\c{c}}alves, Vasco and Vieira, Pedro",
    title = "{Light-Cone Bootstrap of Higher Point Functions and Wilson Loop Duality}",
    eprint = "2008.10407",
    archivePrefix = "arXiv",
    primaryClass = "hep-th",
    doi = "10.1103/PhysRevLett.126.121603",
    journal = "Phys. Rev. Lett.",
    volume = "126",
    number = "12",
    pages = "121603",
    year = "2021"
}

@article{Toller:1969gx,
    author = "Toller, M.",
    title = "{On some properties of the multi-particle amplitude expressed as a function of group-theoretical variables}",
    doi = "10.1007/BF02731814",
    journal = "Nuovo Cim. A",
    volume = "62",
    pages = "341--371",
    year = "1969"
}

@article{Bern:2011qt,
    author = "Bern, Zvi and Huang, Yu-tin",
    title = "{Basics of Generalized Unitarity}",
    eprint = "1103.1869",
    archivePrefix = "arXiv",
    primaryClass = "hep-th",
    reportNumber = "UCLA-11-TEP-103",
    doi = "10.1088/1751-8113/44/45/454003",
    journal = "J. Phys. A",
    volume = "44",
    pages = "454003",
    year = "2011"
}

@article{Boels:2012if,
    author = "Boels, Rutger H.",
    title = "{Three particle superstring amplitudes with massive legs}",
    eprint = "1201.2655",
    archivePrefix = "arXiv",
    primaryClass = "hep-th",
    doi = "10.1007/JHEP06(2012)026",
    journal = "JHEP",
    volume = "06",
    pages = "026",
    year = "2012"
}

@article{Boels:2012ie,
    author = "Boels, Rutger H. and O'Connell, Donal",
    title = "{Simple superamplitudes in higher dimensions}",
    eprint = "1201.2653",
    archivePrefix = "arXiv",
    primaryClass = "hep-th",
    doi = "10.1007/JHEP06(2012)163",
    journal = "JHEP",
    volume = "06",
    pages = "163",
    year = "2012"
}

@article{Chakraborty:2020rxf,
    author = "Chakraborty, Soumangsu and Chowdhury, Subham Dutta and Gopalka, Tushar and Kundu, Suman and Minwalla, Shiraz and Mishra, Amiya",
    title = "{Classification of all 3 particle S-matrices quadratic in photons or gravitons}",
    eprint = "2001.07117",
    archivePrefix = "arXiv",
    primaryClass = "hep-th",
    reportNumber = "TIFR/TH/20-1",
    doi = "10.1007/JHEP04(2020)110",
    journal = "JHEP",
    volume = "04",
    pages = "110",
    year = "2020"
}

@article{Caron-Huot:2022jli,
    author = "Caron-Huot, Simon and Li, Yue-Zhou and Parra-Martinez, Julio and Simmons-Duffin, David",
    title = "{Graviton partial waves and causality in higher dimensions}",
    eprint = "2205.01495",
    archivePrefix = "arXiv",
    primaryClass = "hep-th",
    reportNumber = "CALT-TH 2022-17",
    doi = "10.1103/PhysRevD.108.026007",
    journal = "Phys. Rev. D",
    volume = "108",
    number = "2",
    pages = "026007",
    year = "2023"
}
\end{document}